\documentclass[12pt]{iopart}
\usepackage{epsf} 
\usepackage{epsfig} 
\usepackage{amssymb} 
 
\begin{document} 
 
\title[Growing dynamical length, scaling and heterogeneities
in the 3D {\sc EA} model]
{Growing dynamical length, scaling and heterogeneities
in the 3D Edwards-Anderson model} 
\vskip 10pt 
\author{Ludovic D. C. Jaubert$^{1,2}$, Claudio Chamon$^3$,  
Leticia F. Cugliandolo$^2$ and Marco Picco$^2$ 
} 
\address{ 
$^1$ Laboratoire de Physique,  
Ecole Normale Sup\'erieure de Lyon, France \\ 
$^2$ Universit\'e Pierre et Marie Curie -- Paris VI,  
Laboratoire de Physique Th\'eorique et Hautes Energies, 
Jussieu, France\\ 
$^3$ Physics Department, Boston University, USA \\ 
} 
\date{today} 
 
\begin{abstract} 
We study numerically spatio-temporal fluctuations during the
out-of-equilibrium relaxation of the three-dimensional
Edwards-Anderson model. We focus on two issues. (1) The evolution of a
growing dynamical length scale in the glassy phase of the model, and
the consequent collapse of the distribution of local coarse-grained
correlations measured at different pairs of times on a single function
using {\it two} scaling parameters, the value of the global
correlation at the measuring times and the ratio of the coarse
graining length to the dynamical length scale (in the thermodynamic
limit).  (2) The `triangular' relation between coarse-grained local
correlations at three pairs of times taken from the ordered instants
$t_3 \leq t_2 \leq t_1$.  Property (1) is consistent with the
conjecture that the development of time-reparametrization invariance
asymptotically is responsible for the main dynamic fluctuations in
aging glassy systems as well as with other mechanisms proposed in the
literature. Property (2), we stress, is a much stronger test of the
relevance of the time-reparametrization invariance scenario.
\end{abstract} 
 
\section{Introduction} 
 
Studies of structural glasses have provided mounting evidence, both
from numerical 
simulations~\cite{Laird}-\cite{Ludovic}
and experimental 
probes~\cite{Ediger}-\cite{Berthier-etal},
that there is a growing length scale as a glass former is
supercooled and the dynamics is heterogeneous across the system,
with cooperative particle rearrangements occurring at the nanometer
scale. More recent analysis in the glassy phase also suggests the
existence of a growing length-scale and a heterogeneous character of
the dynamics~\cite{Vollmayr}-\cite{Castillo2}.
These findings reinforce the need to investigate the
spatio-temporal fluctuations of glassy systems as a possible route to
better understand the origin of the dramatic dynamical slow down in
the glassy phase.

The same two characteristics, a growing dynamical length scale and 
spatial heterogeneities, also occur in the possibly better understood 
spin glass problem. Indeed, while a precise and unambiguous connection 
between problems with quenched disorder and structural glasses
is still lacking, the possibility that such connection exists has been 
put forward dating back to the work by Kirkpatrick, Thirumalai and 
Wolynes~\cite{KTW1}-\cite{KTW3}, who argued for a connection between 
structural fragile glasses and the $p$-spin disordered model. More 
recently, Tarzia and Moore~\cite{Tarzia-Moore} argued that, in 
the presence of a uniform field, the Edwards-Anderson model, even in 
one dimension, shares many features with the structural glass problem, 
such as an apparent Kauzmann temperature and a Vogel-Fulcher relaxation 
law. Hence, understanding the spatio-temporal fluctuations in the 
Edwards-Anderson model, in addition to being interesting in itself, 
may help elucidate similar issues in structural glasses if either a 
connection indeed exists between these problems, or heterogeneous 
dynamics is universal in glasses even though the mechanism for the slow 
relaxation may differ in those two systems. 
 
In magnetic systems, 
aging can be globally quantified by the two-time spin-spin 
correlations, summed over all sites in the system.  One of the 
consequences of heterogeneous dynamics in spin glasses is that aging, 
a manifestation of the breakdown of time translation invariance, 
should be non-uniform across the system. The random and  
quenched nature of the interactions naturally introduces in these cases  
a different dynamics from site to site~\cite{Federico}-\cite{Ca03}.  
In connection to the structural  
glass problem one is, though, interested in quantifying fluctuations  
that are not dictated by the random interactions but that is inherent to the  
glassy dynamics.   
Such kind of local dynamics and, in particular, local 
aging~\cite{Ca03} can be described by the two-time spin-spin 
correlations, which instead of being spatially averaged over the whole 
bulk volume, are only averaged over a coarse-graining cell with volume 
$V_r=\ell^3$ centered at some site $r$: 
\begin{equation} 
C_r(t,t_w) = \frac{1}{\ell^3} \sum_{i\in V_r} s_i(t) s_i(t_w)  
\;. 
\label{eq:Crlocal}
\end{equation} 
The values of such local correlations vary spatially, and can be 
interpreted as representing local and spatially heterogeneous ages
of  the dynamical evolution of the system. 
In this paper we focus on the statistical properties of these two-time
functions. First, we confirm that their spatial fluctuations around the
average serve to define a growing correlation length in the aging
regime. Second, we study the fixed temperature probability
distribution function (pdf) $\rho(C_r;t,t_w,\ell,L)$ for a given pair
of times $t,t_w$ and a given coarse-graining length $\ell$. $L$ is the
linear size of the system. Third, we analyse the triangular relations
between the local coarse-grained two-time functions (\ref{eq:Crlocal})
measured at three pairs of times.  
Let us summarize these studies below before developing them in 
detail in the main text.

We improve the analysis of the pdfs of local correlations presented
elsewhere~\cite{Ca03,Ch04,Castillo1} by taking into account the fact
that the correlation length, and the coarse-graining linear
length, are {\it finite} in simulations and presumably also in
experiments. We show that, indeed, the pdfs of local correlators at
different pairs of times $t,t_w$ can be scaled onto universal curves
as long as the global correlation is the same, {\it and} the ratio of
the coarse graining length over the dynamical correlation length is
held fixed. Such scaling can be understood as follows: at fixed
temperature the pdf $\rho(C_r;t,t_w,\ell,L)$ depends on four
parameters, two times $t,t_w$ and two lenghts $\ell,L$. The dependence
on $t$ and $t_w$ can be replaced by a dependence on $C(t,t_w)$ and
$\xi(t,t_w)$ with the former being the global correlation and the
latter the correlation length evaluated at the measuring times. This
passage is exact. The global correlation is monotonic on the two times
and the correlation length, as we show in
Sect.~\ref{subsect:corr-length}, is a growing function of $t_w$ and a
decreasing function of $C$ thus allowing for the invertion $(t,t_w) \,
\to \,(C,\xi)$. The second step is a scaling assumption: that the pdfs
depend on the coarse graining length $\ell$, the total size $L$, and
the scale $\xi$ only through the ratios $\ell/\xi$ and $\xi/L$. This
last step, we should stress, is really a scaling assumption, and not a
trivial requirement from dimensional analysis, as the lengths $\ell$,
$L$ and $\xi$ are already dimensionless as they are measured in units
of the lattice spacing.  The end result is that the pdfs
characterizing the heterogeneous constant temperature aging of the
system can be written as
\begin{equation} 
\rho[C_r;C(t,t_w), \ell/\xi(t,t_w),\xi(t,t_w)/L] 
\;, 
\end{equation} 
We test this proposal numerically in the $3d$ Edwards-Anderson 
(EA) spin
glass assuming that the thermodynamic limit applies so that the last
scaling ratio disappears.  It is noteworthy that a reasonable scaling
with the global correlation held fixed and ignoring the correlation
length dependence has been already achieved, {\it approximately}, in
the $3d$EA model~\cite{Ca03}, as well as in kinetically
constrained models~ \cite{Ch04} and Lennard-Jones
systems~\cite{Castillo1}. Our finding that one has to hold the ratio
$\ell/\xi$ constant for a full collapse, as shown here for the $3d$EA 
model, should also be applied in these last two
cases.  In the random manifold problem one studied the fluctuations of
the global two-time roughness of a finite elastic line (finite $L$)
thus reducing the two parameters, $\ell/\xi$ and $L/\xi$, to the same
one~\cite{elastic2}. The importance of having to take into account the
growing correlation length in the scaling analysis of the dynamics of
super-cooled liquids was stressed by Berthier~\cite{Ludovic}.

Next, we analyse the relation between local correlations measured at
three pairs of times $(t_3,t_2)$, $(t_2,t_1)$ and $(t_3,t_1)$ on the
same spatial point. We argue that the comparison between the local
triangular relation and the global one can be used as a very stringent
test of the time-reparametrization invariance scenario for the origin
of dynamic fluctuations in glassy dynamics~\cite{Ca03,Ch04}. 

The paper is organized as follows. In Sect.~\ref{sec:themodel} we
present the definition of the $3d$EA model.
Section~\ref{sec:global-corr} recalls the main properties of the decay
of the global correlation function. We discuss here several scaling
forms proposed in the literature and we adopt the one that gives 
the best description of the data in the numerical time-window. We also
recall the parametric construction relating the values of the global
correlations at three different pairs of times that we shall use later
in the local context. In Sect.~\ref{sec:local} we discuss the
statistics of the local coarse-grained two-time correlation
functions. We analyse the two-time dependent growing correlation
length and we present its most convenient representation as a function
of $t_w$ and the value of the global correlation.  We study the pdf
of local coarse-grained correlations and their dependence on the
coarse-graining length $\ell$ and times $t$ and $t_w$. We analyse the
triangular relations between local coarse-grained correlations and we
explain why the time-reparametrization invariance scenario makes a
very concrete prediction for their behaviour.  Finally, in
Sect.~\ref{sec:conclusions} we present our conclusions.

\section{The model} 
\label{sec:themodel} 
 
We focused on the three-dimensional Edwards-Anderson ($3d$EA)  
spin-glass model defined by the energy 
\begin{equation} 
H = \sum_{\langle \, ij\, \rangle} J_{ij} s_i s_j 
\; .  
\end{equation} 
The interaction strengths $J_{ij}$ act on nearest neighbours on a 
cubic $3d$ lattice and are independent identically 
distributed random variables taken from a Gaussian probability 
distribution with zero mean and unit variance.  
The spins $s_i$ 
with $i=1,\dots, N=L^3$ are bimodal Ising variables. The static  
critical temperature is $T_g\sim 0.92$~\cite{Pl05}.  
 
We simulated an instantaneous quench from infinite temperature by 
choosing a random initial condition $s_i(0)=\pm 1$ with probability a 
half. We used Monte Carlo dynamics with the Metropolis algorithm to 
mimic the temporal relaxation of the spin-glass coupled to an external 
environment at temperature $T$. A Monte Carlo step (MCs) corresponds to 
$N$ randomly chosen spin flip attempts. 
 
We used lattices with linear size $L=60$ and  $L=100$ and, in order to reduce
finite size effects, we set periodic boundary conditions in all
directions.  We present simulations at $T=0.6$; this value is optimal
in the sense that it is not too close to the critical point and
critical fluctuations should be thus reduced, and it is not too low so
that the dynamics is not prohibitively sluggish.  Similar results are
obtained for a bimodal distribution of coupling strengths.  For the
time-scales reached, finite size effects are not important, as we
checked by varying the system sizes.

\section{Global self-correlation} 
\label{sec:global-corr} 
 
The two-time {\it self-correlation} 
\begin{equation} 
C(t,t_w) = N^{-1} \sum_{i=1}^N s_i(t) s_i(t_w) 
\;  
\label{eq:self-correlation}  
\end{equation} 
measures the overlap of the spin configurations at times $t_w$, the 
waiting-time, and $t$, the measuring time. It characterizes the {\it 
global} or {\it macroscopic} non-equilibrium relaxation of glassy 
systems. In the large $N$ limit and for long but finite times, that is 
to say well before equilibration, the 
self-correlation~(\ref{eq:self-correlation}) is self-averaging with 
respect to noise and disorder induced fluctuations~\cite{LFC-houches}. 
By using then a  few samples we reach longer time scales ($\sim 10^8$ MCs) 
for similar system sizes   
than previously done in the literature.  
 
The two-time dependence of (\ref{eq:self-correlation}) is 
well-understood in coarsening systems~\cite{AB94}  
away from criticality where two two-time regimes  
are simply ascribed to equilibrium thermal fluctuations 
within the ordered domains and non-equilibrium domain-wall motion. In 
the limit $t\geq t_w \gg t_0$, with $t_0$ some microscopic time scale, 
the correlation can be written as a sum of two terms representing 
these two-time regimes: 
\begin{eqnarray} 
C(t,t_w) = C_{st}(t-t_w) + C_{ag}(t,t_w)  
\label{eq:additive-scaling} 
\end{eqnarray} 
with the limit conditions 
\begin{eqnarray} 
\begin{array}{ll} 
C_{st}(0) = 1-q_{ea} \;,  
\qquad  
&  
\lim_{t_w\to t^-}  
C_{ag}(t,t_w) =q_{ea} 
\; ,  
\nonumber\\ 
\lim_{t-t_w \to \infty} C_{st}(t-t_w) = 0 \; ,  
\qquad  
& 
\lim_{t\gg t_w}  
C_{ag}(t,t_w) =0  
\; .  
\end{array} 
\end{eqnarray} 
$q_{ea}$ is a measure of the equilibrium ordering.  A finite limit of 
the equilibrium -- stationary -- part of the correlation, $C_{st}$, is 
a {\it sine qua non} condition for the existence of a finite 
temperature phase transition. 
 
In ferromagnets and $p$ spin disordered models the 
non-equilibrium term $C_{ag}$ satisfies dynamic scaling, 
\begin{equation} 
C_{ag}(t,t_w) \sim q_{ea} \;  
f\left(\frac{R(T,t)}{R(T,t_w)} \right)  
\; ,  
\label{eq:dynamic-scaling} 
\end{equation} 
with $R(T,t)$ a monotonic function of time.
The label $_{ag}$ indicates the aging character 
of the dynamics in the sense that older systems (longer $t_w$'s) have 
a slower decay than younger ones (shorter $t_w$'s).  Dynamic scaling 
(\ref{eq:dynamic-scaling}) implies that any three correlations  
between three generic times $t_0 \ll t_3 < t_2 < t_1$ and taking  
values below 
the plateau at $q_{ea}$ are related by 
\begin{equation} 
C_{ag}(t_1,t_3) = q_{ea} \; f\left\{  
f^{-1}[C_{ag}(t_1,t_2)/q_{ea}] \;    
f^{-1}[C_{ag}(t_2,t_3)/q_{ea}] 
\right\} 
\; ,  
\label{eq:triangular} 
\end{equation}  
independently of the functional form of $R(T,t)$.

In the Sherrington-Kirkpatrick
(SK) model~\cite{LFC94} the aging part of the correlation has itself a 
hierarchical organisation and satisfies dynamic ultrametricity 
\begin{equation} 
C_{ag}(t_1,t_3) = \min[C_{ag}(t_1,t_2), C_{ag}(t_2,t_3)] 
\label{eq:dyn-ultra} 
\end{equation} 
for generic times $t_1 > t_2 > t_3 \gg t_0$ such that  
all correlations take values below the plateau at $q_{ea}$.   
 The asymptotic character of this result 
makes it  difficult to be observed  in a numeric 
simulation. A preasymptotic scaling  
\begin{equation} 
C_{ag}(t,t_w) \sim q_{ea} \;  
f\left(\frac{R(T_c,t-t_w)}{R(T_c,t_w)} \right)  
\; ,  
\qquad \mbox{with} \qquad  
R(T_c,t) \sim \ln t/t_0 
\; . 
\label{eq:dynamic-scaling-critical-trap}  
\end{equation} 
that reaches the ultrametric relation (\ref{eq:dyn-ultra}) in the long
times limit was found in the trap model at its critical
temperature~\cite{Be02}.
 
Several numerical studies of the self-correlation
(\ref{eq:self-correlation}) in the $3d$EA model appeared in the
literature~\cite{Anderson}-\cite{Berthier02}.  The first difficulty
encountered is the identification of a plateau separating equilibrium
and out-of-equilibrium relaxations.  This separation is not as clear
as desired in the numerical data that could as well be described with
a scaling that is compatible with the absence of a finite temperature
phase transition -- a multiplicative scaling as found for instance in
problems of elastic lines in low dimensional random
environments~\cite{elastic2,elastic}. However, evidence for a finite
temperature phase transition in the $3d$EA model~\cite{Caracciolo}
suggests the additive scaling~(\ref{eq:additive-scaling}).  Experiments
on an insulating spin-glass sample are consistent with this type of
scaling too~\cite{Herisson}.
 
\begin{figure}[h] 
\centerline{ 
\includegraphics[width=10cm]{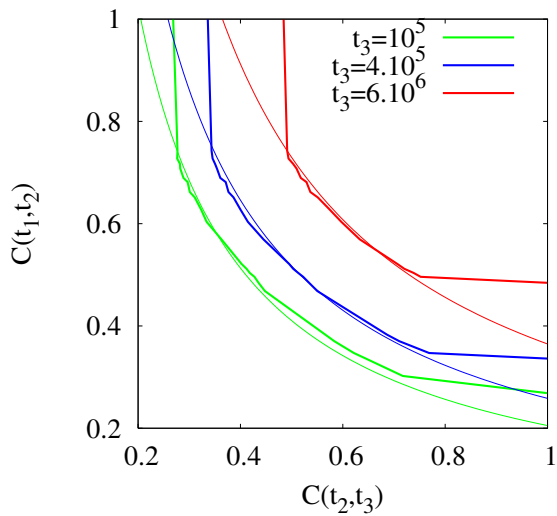} 
} 
\caption{Parametric plot of the correlation functions $C(t_1,t_2)$
against $C(t_2,t_3)$ using $t_2$ as a parameter varying between $t_3$
and $t_1$. The curves are averaged over 5 or 7 samples. 
We use $t_1=1.3 \times 10^8$ MCs and three values of $t_3$
(given in the key) that 
lead to three values of $C(t_1,t_3)$.  As a
guide-to-the-eye we show with thin solid lines the hyperbolic
functions $q_{ea}\; C(t_1,t_3)/C(t_2,t_3)$ that fit the data correctly
in the curved parts of the plot.}
\label{fig:triangular} 
\end{figure}

The existence of a special value of $C$ at $q_{ea}$ as well as the
behaviour of the data below this value can be examined by constructing
parametric plots of the correlations and thus testing the relations
(\ref{eq:triangular}) and (\ref{eq:dyn-ultra}).  In
Fig.~\ref{fig:triangular} we plot $C(t_1,t_2)$ against $C(t_2,t_3)$ at
$T=0.6$, using three pairs of $t_3, t_1$ (given in the caption) that
yield different values of $C(t_1,t_3)$.  The construction shows a
rather clear change in regime when $C(t_1,t_2)$ passes through the
value $\sim 0.78$ and when $C(t_2,t_3)$ reaches the same value.  One
can interpret the change from rather straight and flat behaviour
(vertical or horizontal) to the curved piece in-between as a change in
the two-time scaling that operates when each of the correlations goes
beyond $q_{ea}$. (Note that the vertical behaviour when $t_2$ gets
close to $t_1$ is neater than the horizontal behaviour when $t_2$ gets
close to the opposite limit $t_3$; this is a finite time effect,
the two-time scale separation improves at longer times.)  The straight
line pieces indicate an ultrametric relation between correlation
values lying above the breaking point and correlation values lying
below it. This is consistent with an additive separation between a
stationary regime and an aging one.
 
The curved pieces (correlations taking values below the  
breaking points that we identify with $q_{ea}$) 
is very far from a double straight line with a square 
angle as expected for a full dynamic ultrametric behaviour. 
It is instead rather well described by a hyperbolic  
function with parameter $q_{ea} \sim 0.78$. 
The data are not in the 
trully asymptotic regime and before discarding the 
dynamic ultrametric behaviour one should test  
the relevance of its preasymptotic  
form~(\ref{eq:dynamic-scaling-critical-trap}).  

\begin{figure} 
\begin{flushright}
 \includegraphics[width=7.5cm]{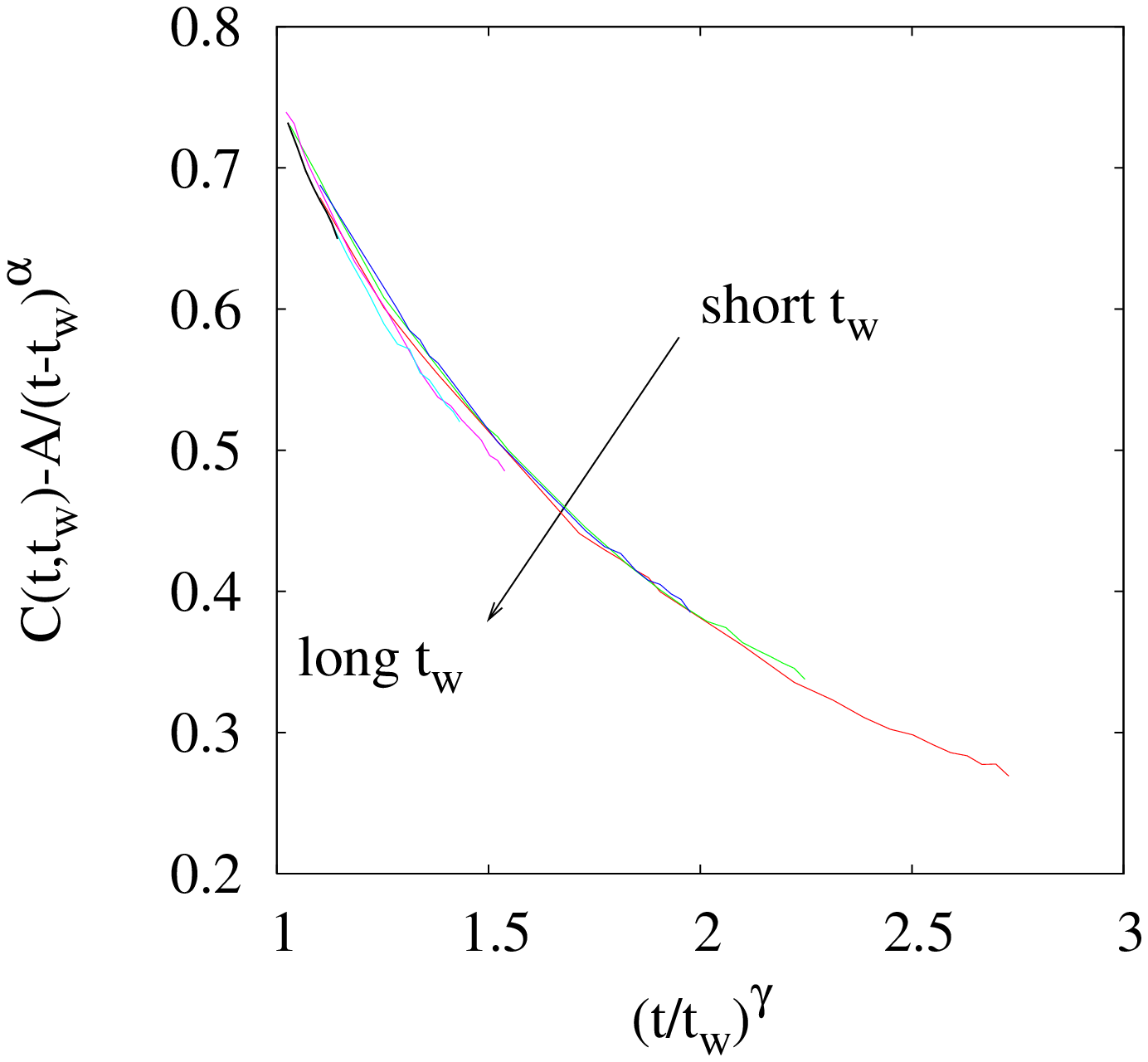} 
 \includegraphics[width=7.5cm]{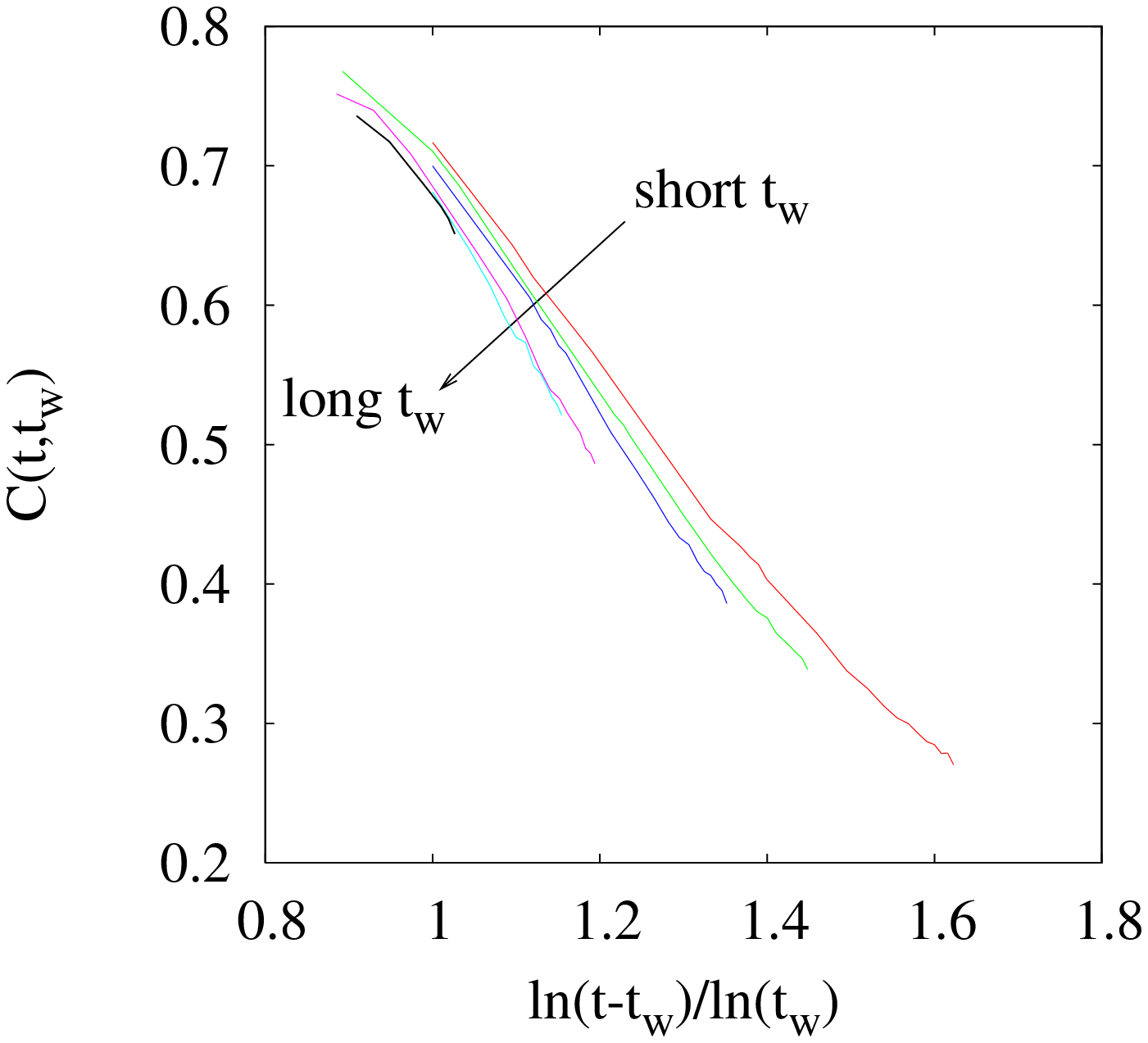} 
\end{flushright} 
\caption{Test of two two-time regimes with simple aging (a), and
ultrametric scaling (b). The waiting-times are $t_w= 10^5, \;
4\times 10^5, \; 10^6, \; 6\times 10^6, \; 10^7$ and $t_w=5\times 10^7$ MCs.
$A=10\pm 2$ and $\alpha=0.5\pm 0.05$ and $\gamma=0.14$.}
\label{fig:scaling-h} 
\end{figure}

In the ideal asymptotic limit of very long-waiting time and very long
time-difference the contribution of the stationary part should
saturate to zero and the full correlation to a constant
$q_{ea}$. However, the approach to the constant can be rather slow --
it is a power in mean-field models -- and thus interfere in the aging
regime. One then
subtracts the stationary contribution, assumed to be a power law,
\begin{equation} 
C_{st}(t-t_w) \sim A (t-t_w)^{-\alpha} 
\end{equation} 
with $A$ and $\alpha$ two fitting parameters~\cite{Vi96}.  A more
refined analysis, including a waiting-time dependent correction to the
stationary part estimated along the lines of the droplet picture,
appeared in \cite{Japan}.  In Fig.~\ref{fig:scaling-h} we compare the
scaling form (\ref{eq:dynamic-scaling}) with $R\sim t^\gamma$, called
`simple aging'~\cite{Vi96} and the pre-asymptotic
ultrametric scaling (\ref{eq:dynamic-scaling-critical-trap}).  The
drift with increasing waiting-time is shown with an arrow. It is clear
that, even though the collapse is not perfect, the simple aging form
yields a much better description of the data. We checked that 
other functions, $R(T,t)$, such as the logarithmic growth of the 
droplet model are not as good. 
   
  In conclusion, even though the numerical data are not utterly
conclusive, they support the existence of a plateau at a finite
$q_{ea}$, with a rather slow approach of the stationary part to it,
and an aging contribution that is best described by a simple aging
form, $R(t) \sim t^\gamma$ with $\gamma\sim 0.14$. The values of the
other parameters are $q_{ea} \sim 0.8$, $A=10$ and $\alpha\sim 0.5$;
the function $f(x)$ is then very close to $f(x) \sim q_{ea}/x$. In 
short, 
\begin{equation}
C_{st}(t-t_w) \sim 10 \; (t-t_w)^{-0.5} 
\; , 
\qquad 
C_{ag} \sim 0.8 \, \left( \frac{t_w}{t} \right)^{0.14}
\; . 
\label{eq:scaling-C3dEA}
\end{equation}
Note that this form is totally different from the analytic solution to
the SK model~\cite{LFC94}, the mean-field version of the $3d$EA
model. We shall use the triangular construction in
Sect.~\ref{subsec:triangular} to analyse dynamic fluctuations.

\section{Local fluctuations} 
\label{sec:local} 
 
The local coarse-grained correlation is defined similarly to the  
global one~\cite{Ca03} 
\begin{equation} 
C_r(t,t_w) = N_r^{-1} \sum_{i=1}^{N_r} s_i(t) s_i(t_w)  
\end{equation} 
where the sum runs over the $N_r=\ell^3$ spins lying in a cubic box
with linear size $\ell$ centered at the site $r$. We work with
overlapping coarse-graining boxes centered on each site of the
lattice. In this way we improve the statistics obtaining $N$ data
points {\it per} sample. It is important to notice that working with
Ising spins we have a natural discretization of the local
correlations: $C_r$ varies between $-1$ and $1$ in steps of
$2/\ell^3$. This effect is negligible when $\ell > 6$ in such a way
that $2/\ell^3 < 0.01$ but it is annoying for smaller values, {\it
e.g.} $\ell=3, \; 4$.  Averaging over short time-windows does not
change significantly our results.
 
 
The simplest way to investigate local fluctuations is to study the
probability density function $\rho(C_r)$. This function depends on
five parameters: the two times $t$ and $t_w$, the coarse-graining
length $\ell$, the linear size of the system $L$, and the temperature
$T$ that, in particular, determines the Edwards-Anderson order
parameter $q_{ea}$. In this paper we keep temperature fixed and
henceforth ignore the temperature dependence. We use sufficiently 
large systems sizes so that in practice $L\to\infty$. 

A more refined analysis of the fluctuations that puts the global
time-reparametrization invariance scenario to the test consists in
comparing local correlations measured at three pairs of times and
studying the behaviour of local triangular relations.

\subsection{Evolution of the pdf with the measuring time $t$} 
\label{subsect:generic-features} 
 
\begin{figure} 
\centerline{ 
 \includegraphics[width=7cm]{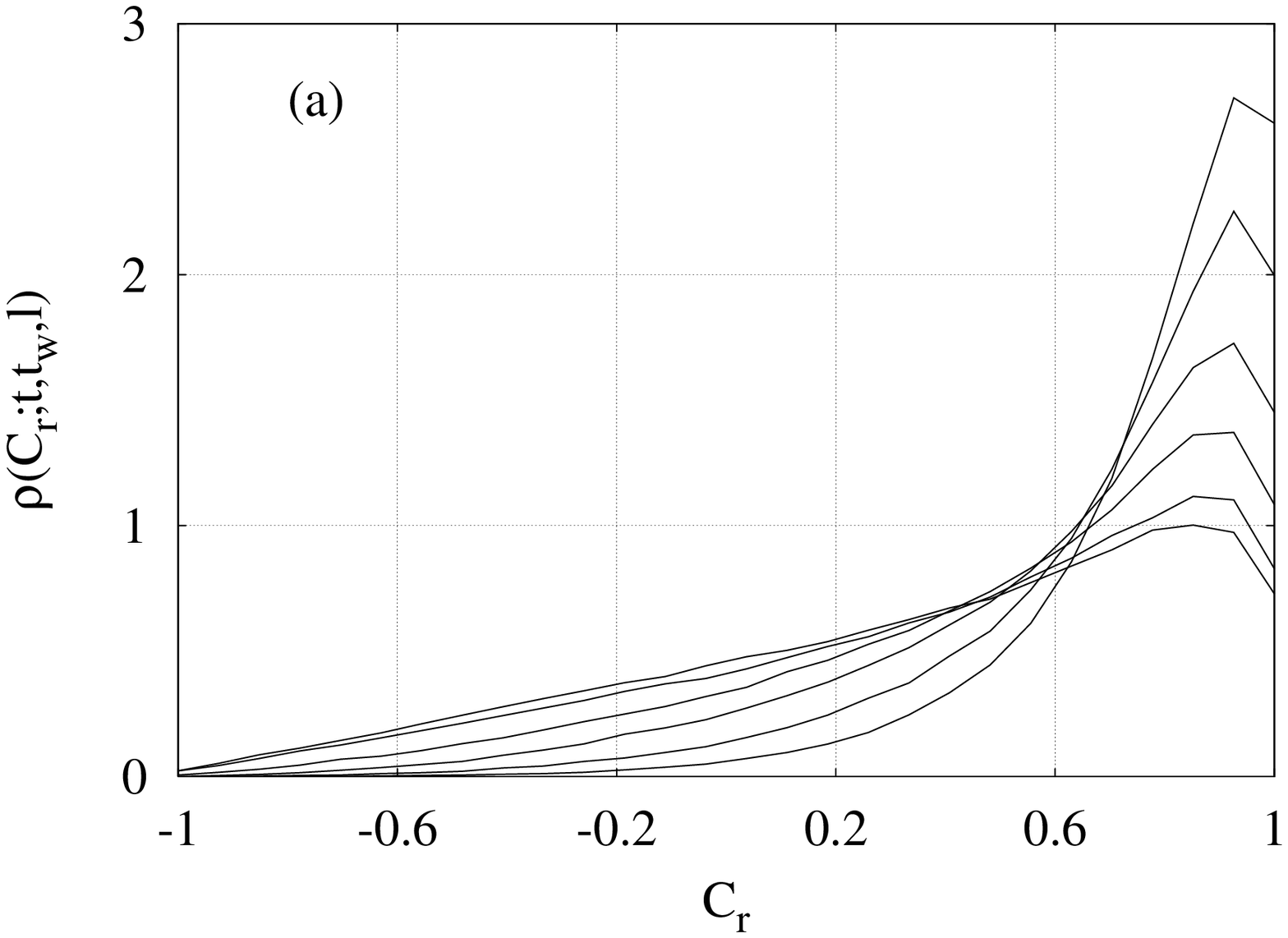} 
\includegraphics[width=7cm]{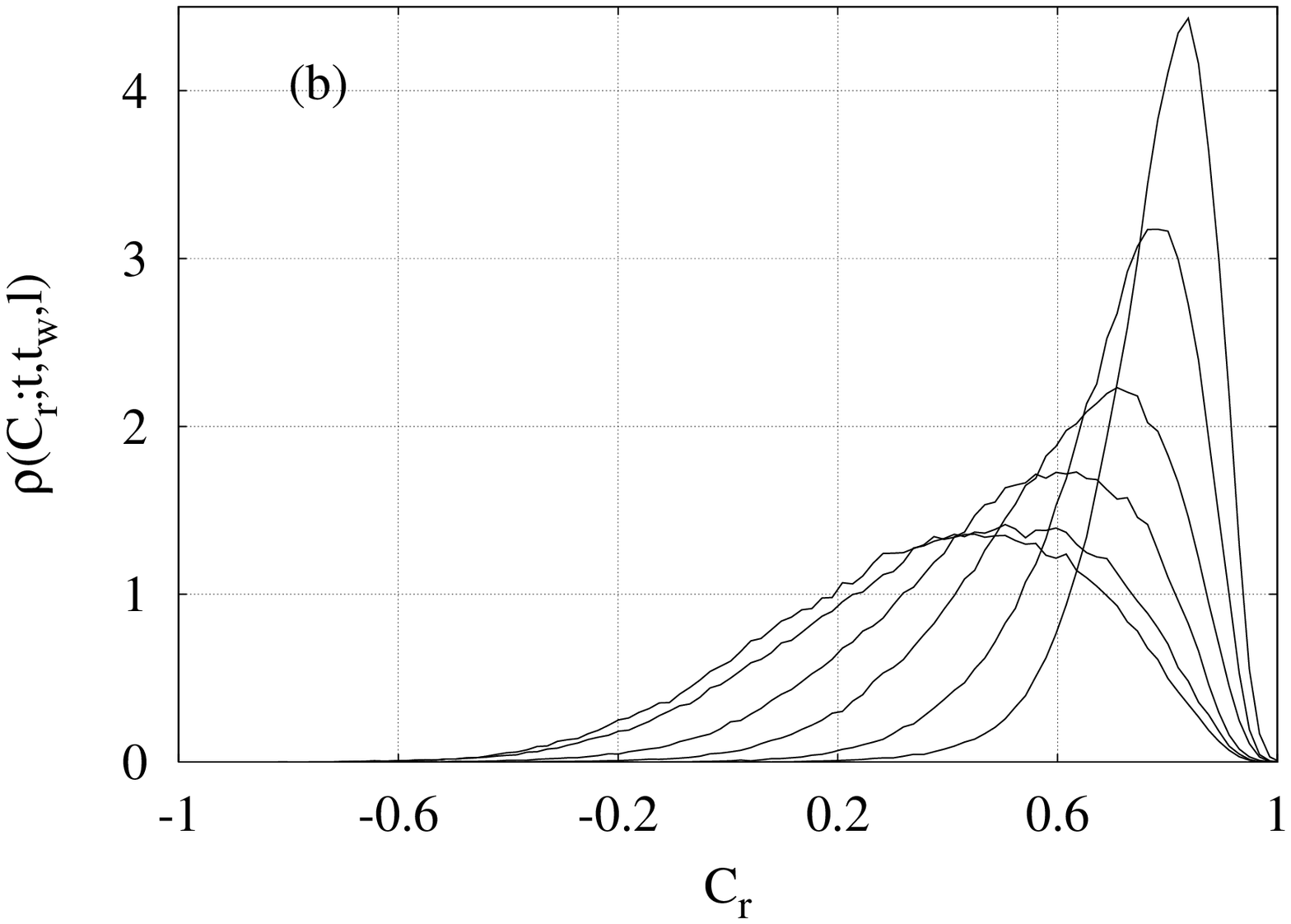} 
} 
\caption{Dependence of the pdf of local coarse-grained correlations on
the value of the global correlation at fixed $t_w$ and varying
$t$. The waiting time is fixed to $t_w= 5\times 10^4$ MCs. The total
times are $t=6.5 \times 10^4$ MCs ($C=0.78$), $t=1.2 \times 10^5$
MCs ($C=0.7$), $t=3\times 10^5$ MCs ($C=0.6$), $t=10^6$ MCs ($C=0.5$),
$t=5.5 \times 10^6$ MCs ($C=0.4$), and $t=9 \times 10^6$ MCs
($C=0.36$), and leads to the global correlation values given between
parenthesis. Coarse-graining box with linear size $\ell=3$ (a) and
$\ell=6$ (b).  The height of the peak diminishes and stays located at
the same $C$ in (a) while it drifts to the left in (b). The left tail
grows with increasing $t$ in both cases.}
\label{fig:rhoCr-generic} 
\end{figure}

We start by analysing the evolution of $\rho$ with the value of its 
mean, or global self-correlation, $C$, obtained using $t_w= 5 \times 
10^4$ MCs and the corresponding $t$'s (given in the caption) at fixed 
and rather small coarse-graining length, $\ell=3$. The curves in 
Fig.~\ref{fig:rhoCr-generic}~(a) show that for all $C$ values there is 
a peak at $C_r\sim 0.85$ that is slightly higher than the value of 
$q_{ea}$ estimated from the study of the global self-correlation.  The 
height of the peak diminishes with decreasing $C$ while the weight on 
the tail on negative $C_r$'s increases.
Using such a small coarse-graining volume, many boxes partially reverse 
when times are sufficiently separated.  Two undesired effects of using 
coarse-graining volumes that are too small are thus clearly identified 
and can be easily interpreted within a domain-growth picture of two 
competing equilibrium states. At long waiting-times a patchwork of 
domains with relatively large radii establishes. The density of domain 
walls is small. Using a small coarse-graining volume one has a high 
probability of contouring a region that remains in the `basin of 
attraction of the equilibrium state' in which it is at time $t_w$, 
with no domain-walls crossing the box -- independently of how many of 
these states exist. In these cases, the local coarse-grained 
correlation is reduced from one just by thermal fluctuations and 
should then be very close to $q_{ea}$ (if the size of the 
coarse-graining box is really small one can imagine that not even all 
the thermal fluctuations needed to reach $q_{ea}$ enter the box and 
thus the peak can be at a higher value).  When time passes, more and 
more of these boxes are crossed or traversed by domain walls and the 
weight of the peak diminishes while the tail of the pdf increases.  In 
a simple coarsening picture with only two equilibrium states that 
develop relatively rapidly and thin domain walls between them, one 
should see a second peak developing at $-q_{ea}$ due to regions that 
have reversed, from one equilibrium state to the other, after the 
passage of a domain-wall (or an odd number of them). For the values of 
the global correlation (total times) reached we do not see this 
process here. 
 
The stationary character of the peak at $C_r\sim q_{ea}$ is lost 
using a larger coarse-graining box, as shown in 
Fig.~\ref{fig:rhoCr-generic}~(b). This panel also shows the evolution with 
$t$ of the pdf of local correlations at fixed $\ell=6$. When $t=t_w$ 
the pdf is just a delta function at one. As $t$ departs from $t_w$ one 
enters the stationary regime and the pdf is approximately Gaussian 
around its mean value $C> q_{ea}$. Upon further increasing $t$ one enters the 
interesting aging regime and the pdf departs from being Gaussian, it 
develops a tail towards small values of $C_r$ and its skewness becomes 
negative. At sufficiently long $t$ the pdf starts symmetrizing 
and the fluctuations eventually become Gaussian again when the average 
$C$ reaches zero. 
 
\begin{figure} 
\begin{center}
 \includegraphics[width=7cm]{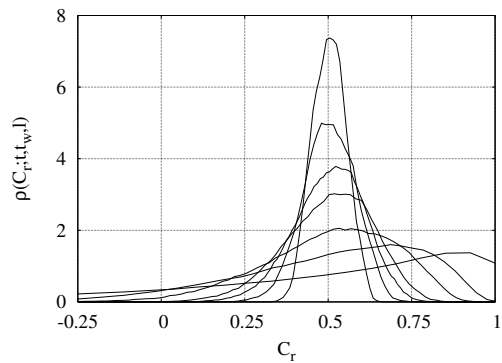} 
\end{center}
\caption{Dependence of the pdf of local coarse-grained correlations on 
the size of the coarse-graining box for fixed total time, $t$, 
and waiting-time, $t_w$; $t_w= 5\times 10^4$ MCs and $t=10^6$ MCs in 
such a way that the global correlation is $C=0.5$ for all data 
sets. The coarse-graining lengths are $\ell=3,\, 5,\, 7,\, 10,\, 12,\, 
15,\, 20$ and as $\ell$ increases the pdfs get narrower and their 
maximum moves towards smaller $C$.} 
\label{fig:rhoCr-generic2} 
\end{figure}

Finally, in Fig.~\ref{fig:rhoCr-generic2} we show the dependence of
the pdf of local coarse-grained correlations on the size of the
coarse-graining box measured at a fixed pair of times $t$ and $t_w$
such that the global correlation is $C=0.5$.  The distribution evolves
from a rather wide negatively skewed form, with the peak at $C_r \sim
q_{ea}$ and an important tail on negative $C_r$'s to a Gaussian pdf
centered at $C$ upon increasing $\ell$.  In the limit $\ell =L$, the
linear size of the system, the distribution concentrates on the global
value $C$.  The figure shows data for $\ell$ varying from $3$ to $20$.
 
It is clear from the analysis above that one needs to use a 
sufficiently large coarse-graining box to avoid finite size and 
discretization effects. To study non-Gaussian statistics, too large 
coarse-graining boxes should be equally avoided. In the next Section we 
analyse the correlation length and we then give a concrete criterion 
to achieve the scaling regime. 

\subsection{The correlation length} 
\label{subsect:corr-length} 
 
\begin{figure} 
\centerline{ 
\includegraphics[width=5cm]{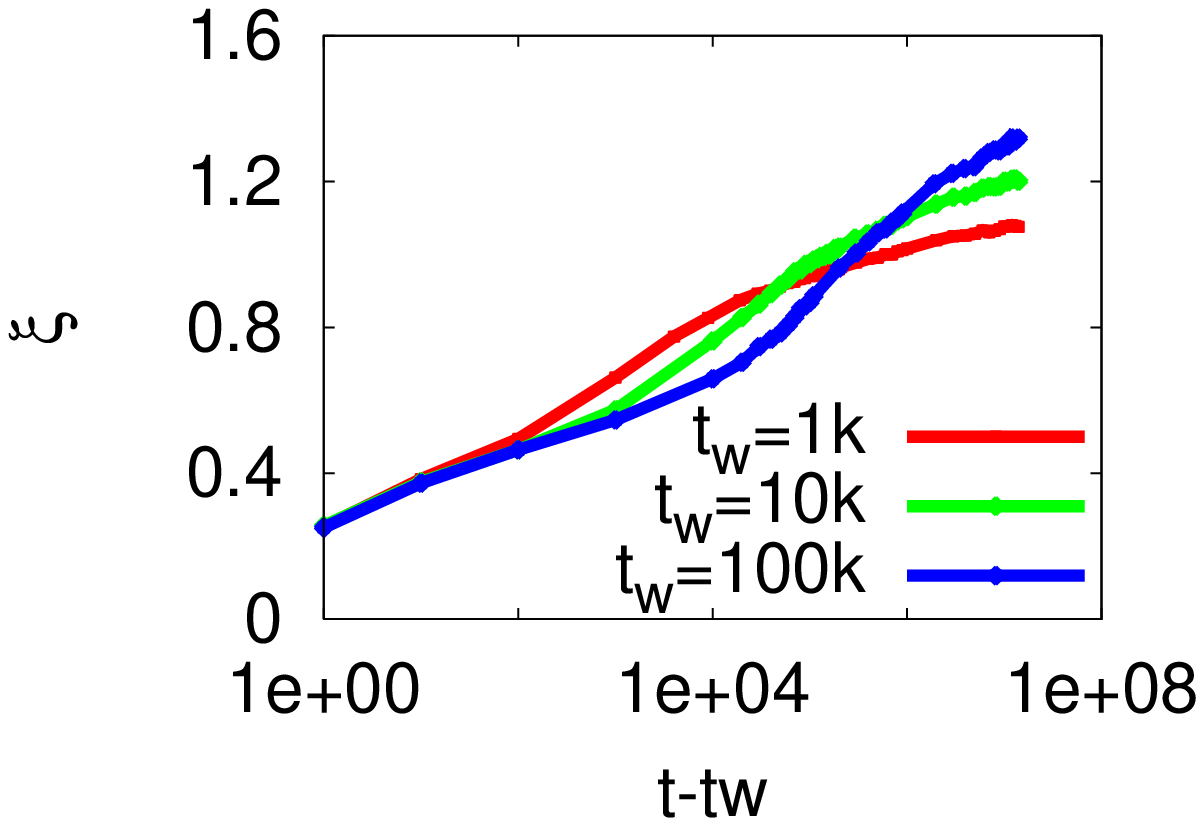} 
\includegraphics[width=5cm]{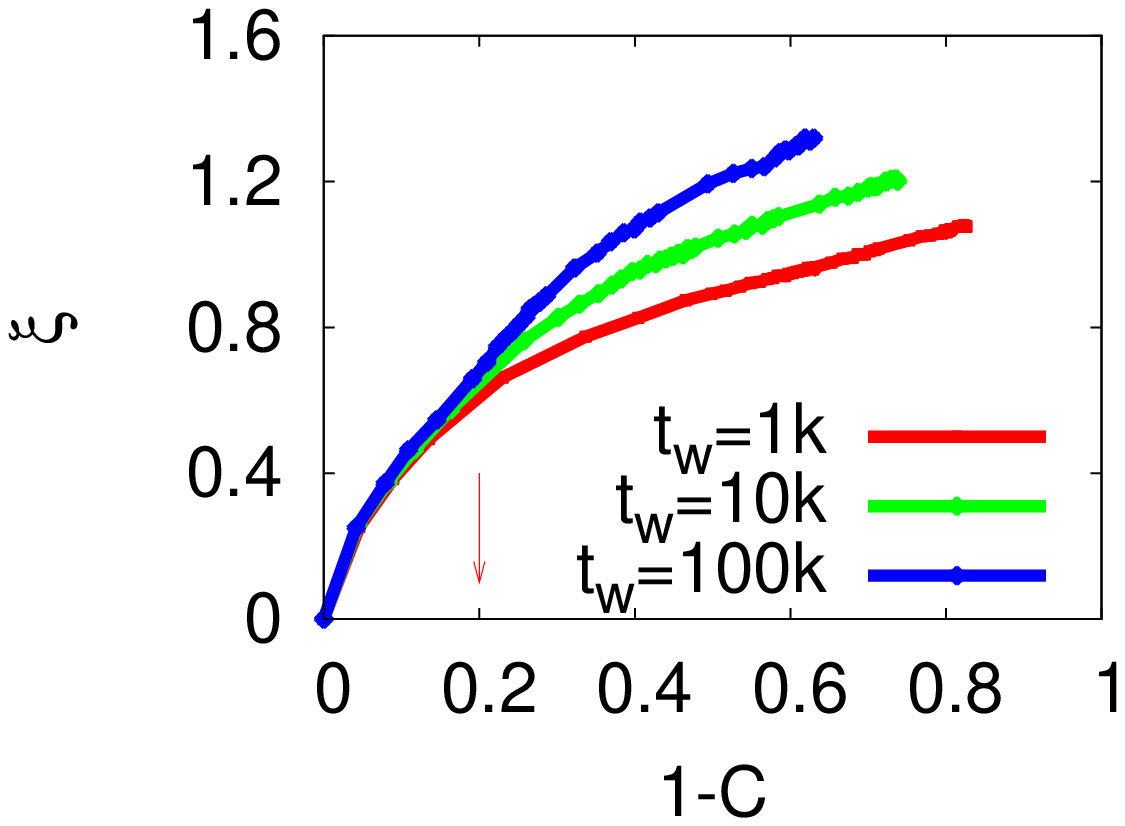} 
\includegraphics[width=5cm]{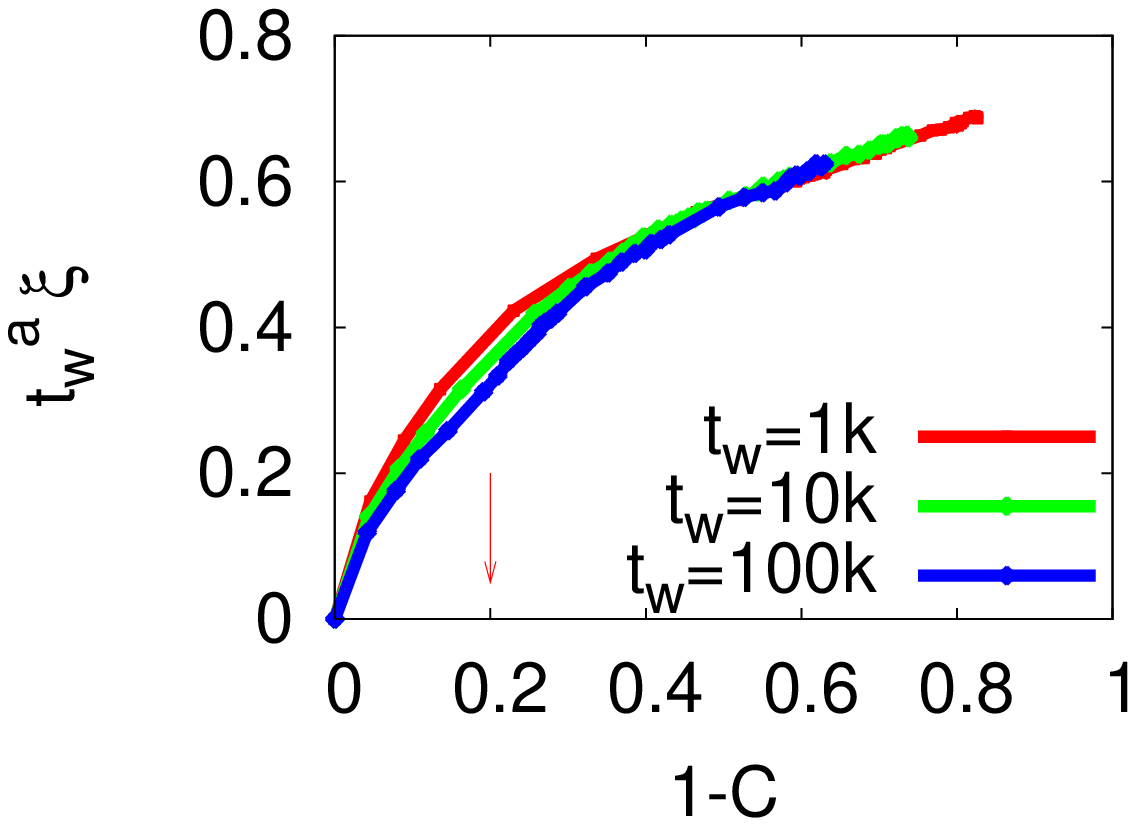} 
} 
\caption{Study of the two-time dependent correlation length
as extracted from the spatial decay of $C_4$ ion a $3d$EA model 
with $L=100$.  (a) As a function of 
time-delay. (b) As a funtion of the global correlation. (c)
Power law scaling with the waiting-time.}
\label{fig:xi} 
\end{figure}

 As in usual ordering processes one would like to identify a
correlation length and determine its temperature and time-dependence.
A {\it two-time} correlation length, $\xi(t,t_w)$,
apt to describe the growth of
order in the low-temperature aging regime, can also be defined from
the study of a two-site two-time correlation. In the spin language one
defines~\cite{Ca03}
\begin{equation} 
C_4(r;t,t_w) = N^{-1} \sum_{i,j=1}^N \; [ \,  
(s_i(t) s_i(t_w) s_j(t) s_j (t_w) -
C^2(t,t_w)\, ]_{|\vec r_i -\vec r_j|=r}  
\;  
\label{eq:Atilde} 
\end{equation} 
and extracts $\xi$ from the fits of its decay as a function of 
$r$ to an exponential. The results of this analysis are
shown in Fig.~\ref{fig:xi}~(a) where we plot $\xi(t,t_w)$ as a function of 
$t-t_w$ for three values of $t_w$ given in the caption. A better 
representation of the same data is given in Fig.~\ref{fig:xi}~(b) where 
we display $\xi$ as a function of $1-C$, evaluated at the same 
times. The curves are now monotonic in both $1-C$   and $t_w$. 
Finally, the $t_w$-dependence can be taken into account 
by proposing, see Fig.~\ref{fig:xi}~(c),  
\begin{eqnarray}
\xi(t,t_w) &\sim&
\left\{
\begin{array}{ll}
\xi_{st}(t-t_w)  &\mbox{for} \;\; C>q_{ea}
\; , 
\\
 t_w^a \; g(C) & \mbox{for} \;\; C<q_{ea}
\; , 
\end{array}
\right.
\label{eq:xi}
\end{eqnarray}
with $a \sim 0.065$ and 
$g(C)$ a monotonically decreasing function of $C$. 

We can also extract the two-time dependence of a typical length scale, to 
be associated to the correlation length, from the analysis of the 
functional form of the pdf of local coarse-grained correlations. 
If the pdfs of 
coarse-grained correlations are close to a Gaussian, one has 
summed over uncorrelated variables and 
the coarse-graining linear length 
has gone beyond the correlation length (at the working times $t$ and 
$t_w$). On the other hand, a pdf that differs from a Gaussian should 
be due to correlations in the underlying variables (see the discussion 
in~\cite{Ch04}). 

 For instance one can analyse the evolution of the skewness $S\equiv
\langle \, (C_r -\langle \, C_r \, \rangle )^3\, \rangle/ \langle \,
(C_r -\langle \, C_r \, \rangle )^2 \, \rangle^{3/2}$ and deduce 
the generic trend of $\xi$ by 
choosing a reference value
$S^*$ to determine the coarse-graining length at which Gaussian
statistics is approached. In this way we obtain $\xi(t_w,C)$ with the same
characteristics as the one shown in Fig.~\ref{fig:xi}.
 
In \cite{Ca03} one found $\xi(t,t_w) \sim (tt_w)^a$ with $a$ a small
exponent or else a logarithmic law of the product of the two times.
We here prefer to use the more precise description in
eq.~(\ref{eq:xi}). Note that the maximum value of the correlation
length reached is just a few lattice spacings, 
also consistent with the results found in the studies using the
two-replica analysis~\cite{Rieger,Berthier02}.

Finally, let us point out a sharp difference between our results for
the dynamical correlation length in the glassy phase of the
$3d$EA model and those found in supercooled liquids. In the
latter, studies of the correlation between the order parameter
fluctuations measured at different times $t$ and $t_w$ on different
regions of a single evolving sample found a non-monotonic behavior as
a function of $\Delta t=t-t_w$~\cite{Donati99,Berthier-etal,Cristina}
(there is no waiting-time dependence there since the system is in a
metastable state with equilibrium-like dynamics). More specifically,
the integral over space of the four point correlation $\chi_4(\Delta
t) \equiv \int d^dr \; C_4(r,\Delta t)$ increases with $Delta t$ until
reaching a maximum at a temperature-dependent $\Delta t^*$ where it
starts falling-off towards zero.  The value of $\Delta t^*$ and the
maximum $\chi_4^*\equiv \chi_4(\Delta t^*)$ increase when approaching
the glass transition, and mode-coupling theories predict their
divergence as a power law of the distance from criticality,
$T-T_{mct}$ at the mode-coupling transition~\cite{MCT}. In our case,
we work not above but, instead, below the glass transition temperature
of the $3d$EA model, and the strictly increasing
correlation length is accompanied by a monotonically increasing
$\chi_4(t-t_w,t_w)$ as a function of $t-t_w$, for fixed $t_w$.

\subsection{Time-scaling} 
\label{subsec:time-scaling}

\begin{figure} 
\centerline{ 
\includegraphics[width=7cm]{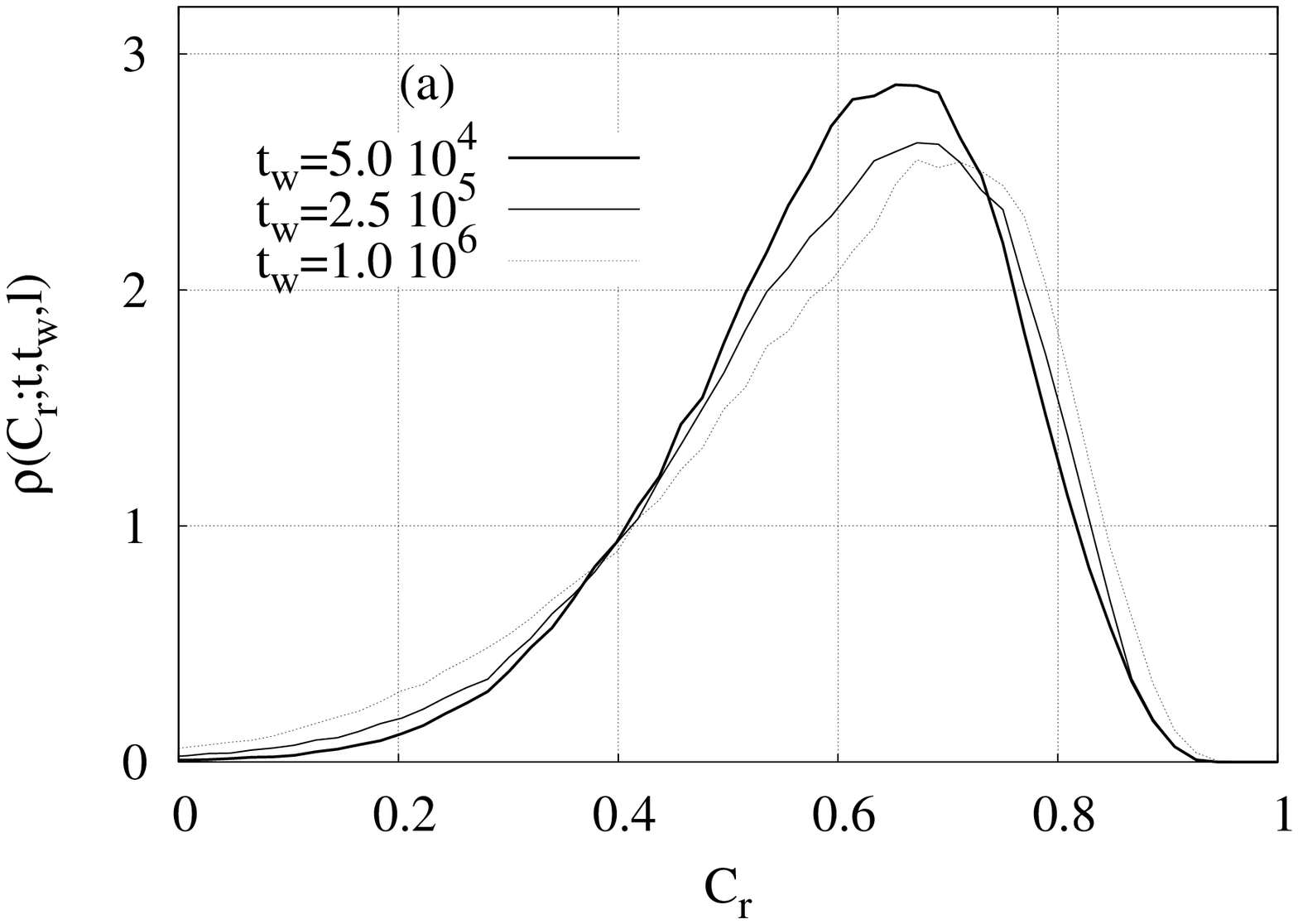} 
\includegraphics[width=7cm]{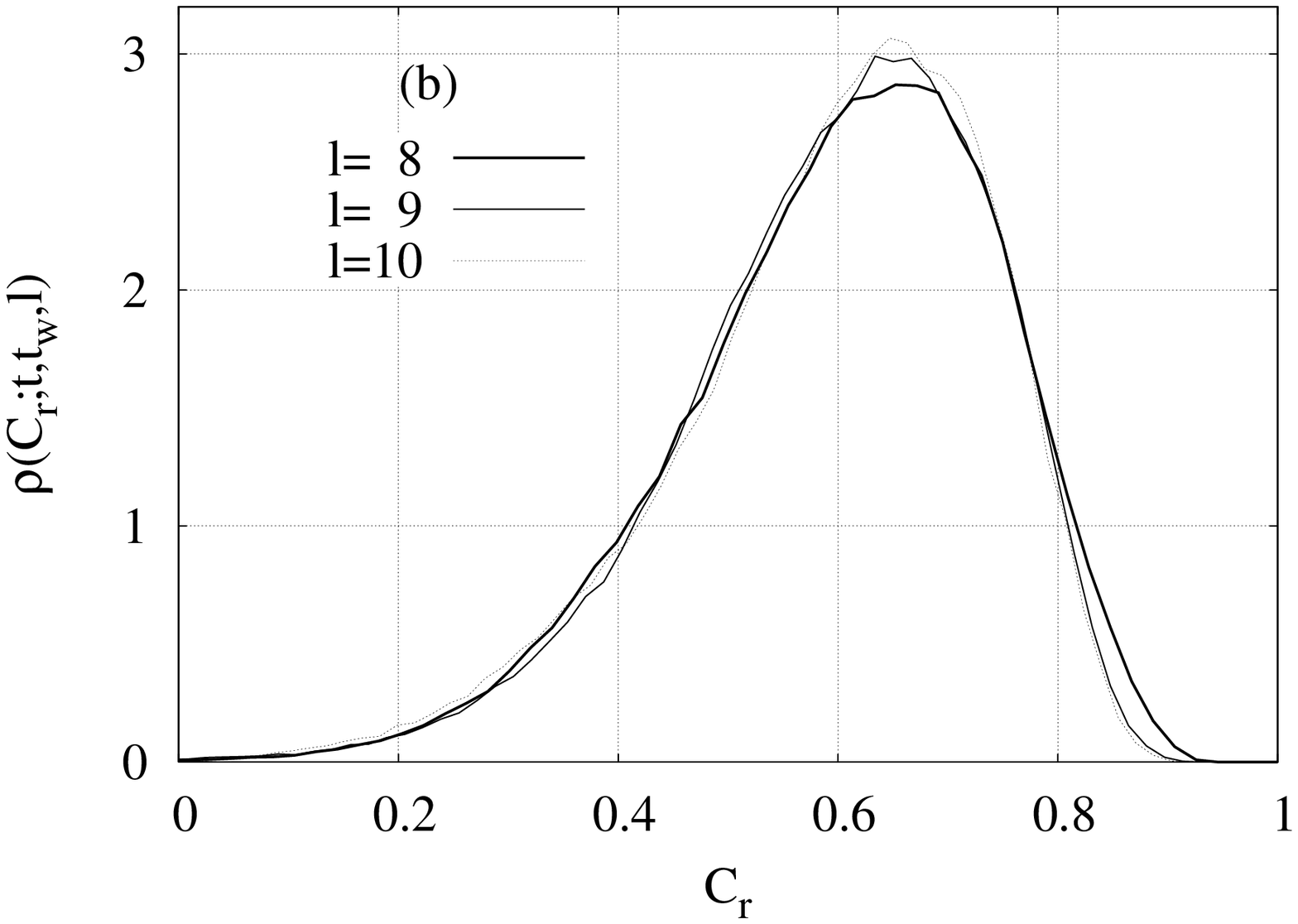} 
} 
\caption{Pdf of local correlations $C_r$. (a) $C_r$ is coarse-grained
on boxes of linear size $\ell=8$ at $C=0.6$ fixed. The curves
correspond to three waiting-times given in the key. There is a slow
drift with increasing $t_w$ (see the text).  (b) $C_r$ is
coarse-grained on boxes with variable length $\ell$. The global
correlation is the same as in panel (a), $C=0.6$. The curves
correspond to $t_w=5 \times 10^4$ and $\ell=8$, $t_w=2.5 \times 10^5$
and $\ell=9$, and $t_w=10^6$ and $\ell=10$.
The collapse is improved with respect to panel (a).} 
\label{fig:time-scaling-pdfs} 
\end{figure}

We are interested in characterizing $\rho(C_r;t,t_w,\ell,L)$ in the
  thermodynamic limit taken at the outset ($L\to\infty$), and in the
  asymptotic limit taken with a prescription that we now explain.
  Exploiting once again the monotonic character of the two-time global
  correlation, the probability density becomes $\rho(C_r; C,
  t_w,\ell)$ where we keep the name $\rho$ to simplify the
  notation. Now, we would like to take the limit $t_w\to\infty$ while
  keeping $C$ fixed -- to any value below the plateau $q_{ea}$ -- but
  we need to specify whether we keep $\ell$ finite or we let it go to
  infinity together with $t_w$ to obtain an interesting asymptotic
  limit.  The arguments in \cite{Ca03,Ch04,Ch02} and the analysis of
  the numerical results discussed in Sect.~\ref{subsect:corr-length}
  suggest that $\xi(t,t_w)$ diverges in the limit $t_w\to\infty$
  and/or $t\to\infty$. Thus, if we keep $\ell$ fixed while we let
  $t_w\to\infty$ we are effectively taking smaller and smaller boxes
  with respect to the correlation length $\xi$ and thus considering
  regions that are artificially more and more correlated. This effect
  can be seen by plotting the pdf of local coarse-grained correlations
  at fixed $C$ obtained by using different pairs $t$ and $t_w$ and a
  fixed $\ell$: there is a slow drift of the weight of the pdf towards
  the peak at $q_{ea}$.  In Fig.~\ref{fig:time-scaling-pdfs}~(a) we
  display the pdf of local correlations $C_r$ for several values of
  the pair of times $t$ and $t_w$ that lead to the same global value
  $C=0.6$. The coarse-graining length is $\ell=5$ in all cases. The
  curves do indeed show a slow drift for increasing $t_w$ with the
  peak moving towards higher values of $C_r$ and an increasing weight
  of the tail on negative values.

We are then forced to  take larger and larger coarse-graining sizes
as $t_w$ and $t$ increase  to attain a stable asymptotic limit of the
probability distributions.  The divergence of $\ell$ should follow
the one of $\xi$ -- that is  relatively slow.  Considering diverging
coarse-graining boxes one  should also get rid of the $C_r$'s that
are close to and above $q_{ea}$  or negative.  Of course, reaching
this limit in a numerical  simulation is impossible and one is forced
to deal with finite times  and finite coarse-graining size effects. 

  The qualitative discussion above can be made quantitative by exploiting
the monotonic character of $\xi$ with $t$ and $C$. We can
actually rewrite the pdf as a function of the global correlation and
the correlation length, $\rho(C_r;C,\xi,\ell)$.  Using now a
natural scaling assumption in which lengths $\xi$ and $\ell$ enter
only through their ratio, then $\rho(C_r; C, \ell/\xi)$.  As in
standard critical phenomena universal behaviour should then appear in
the limit:
\begin{eqnarray} 
\mathop{\lim}\limits_{
\stackrel
{\scriptstyle{(t,t_w,\ell)\to\infty}
}
{
\stackrel{\scriptstyle{C(t,t_w)=C}}
         {\scriptstyle{\ell/\xi\to 0}}
}
}
 \; \; 
\rho(C_r; t,t_w,\ell) = \rho(C_r, C)  
\; . 
\label{limit1} 
\end{eqnarray} 
The limit $\ell/\xi\to 0$ is impossible to impose in a numerical 
simulation since the values of the correlation length are 
relatively small ($\xi <10 $). In practice, we shall then avoid this limit 
and simply work with the requirement 
$\ell/\xi$ fixed
to achieve a good data collapse.
A rough estimate of the growth of the coarse-graining length needed to
scale the data is then given as follows.   The global correlation in
the aging regime scales   as $t/t_w$, approximately (see the analysis
in Sect.~\ref{sec:global-corr}).   In order to keep $C$ constant one
approximately   multiplies the two times $t$ and $t_w$ by the same
constant,   say $\nu$. Under the simultaneous   dilation of times
the correlation length transforms as
\begin{equation} 
\xi(\nu t,\nu t_w) \sim (\nu t_w)^a \; g(C) = \nu^a \; \xi(t,t_w)  
\;  
\end{equation} 
where, for simplicity, we adopted the power-law behaviour 
[see Sect.~\ref{subsect:corr-length} for the analysis of $\xi(t,t_w)$].    
We then estimate the change in coarse-graining length as  
\begin{equation} 
\ell(\nu t,\nu t_w) \sim \nu^{a} \ell(t,t_w) 
\; .  
\label{estimate} 
\end{equation}  
Given that the exponent $a$ is very small, $a\sim 0.1$, the change
in $\ell$ is also quite small. For instance,   if $\nu=2$, the
scaling factor is $\nu^{a} \sim 1.14$.  

In Fig.~\ref{fig:time-scaling-pdfs}~(b) we take into account the 
increasing correlation length with times $t$ and $t_w$ and we then
use  a variable coarse-graining length to study the data at the same
pairs  of times. The values of $\ell$ chosen follow the estimate 
(\ref{estimate}) -- though they slightly differ from it -- and are 
given in the key. With the variable coarse-graining one manages to 
locate the maxima at the same position. The collapse between the two 
longer waiting-times is now very good while the shorter waiting-time 
still has too much weight on large values of $C_r$ and a lower
weight  on the peak than expected.  We believe that this is due to
finite-time  effects. Note also that in the ideal theoretical limit
one should have  no weight at all above $q_{ea} \sim 0.8$ which is
certainly not the  case in the numerical data. The crossover between
stationary and aging  regimes is contaminating our data, especially
at short  waiting-times. Unfortunately, we have reached the limit of
our  numerical time-window and we cannot present results for older
systems.

The analysis of numerical data in other glassy systems should also
take into account the growth of the correlation lenght. In particular,
the approximate but not excellent collapse of the pdfs of local
incoherent correlation functions in Lennard-Jones
mixtures~\cite{Castillo1} should be corrected by taking into account
the need to use a variable coarse-graining length~\cite{Valluzzi}.

\subsection{Triangular relations}
\label{subsec:triangular}

The global
time-reparametrization invariance~\cite{Ca03} suggests that in the ideal
asymptotic limit the slow part of the coarse-grained {\it local}
correlations should scale as
\begin{eqnarray}
&&
C_{ag}(r;t,t_w)  \approx
q_{ea} \; f\left(\frac{h(r,t)}{h(r,t_w)}\right)
\; , 
\label{eq:fluct-cchi}
\end{eqnarray}
with $f$ the {\it same} function describing the global correlation
[eq.~(\ref{eq:dynamic-scaling})] in the ideal limit
$1\ll\ell\ll\xi$~\cite{Ca03}.
The sum rule $C_{ag}(t,t_w) = V^{-1} \int d^dr \; C_{ag}(r;t,t_w)$
applies. We now present some tests of eq.~(\ref{eq:fluct-cchi})
that are based on the parametric representation of the dynamics.

\begin{figure}
\begin{flushright}
\includegraphics[width=8cm]{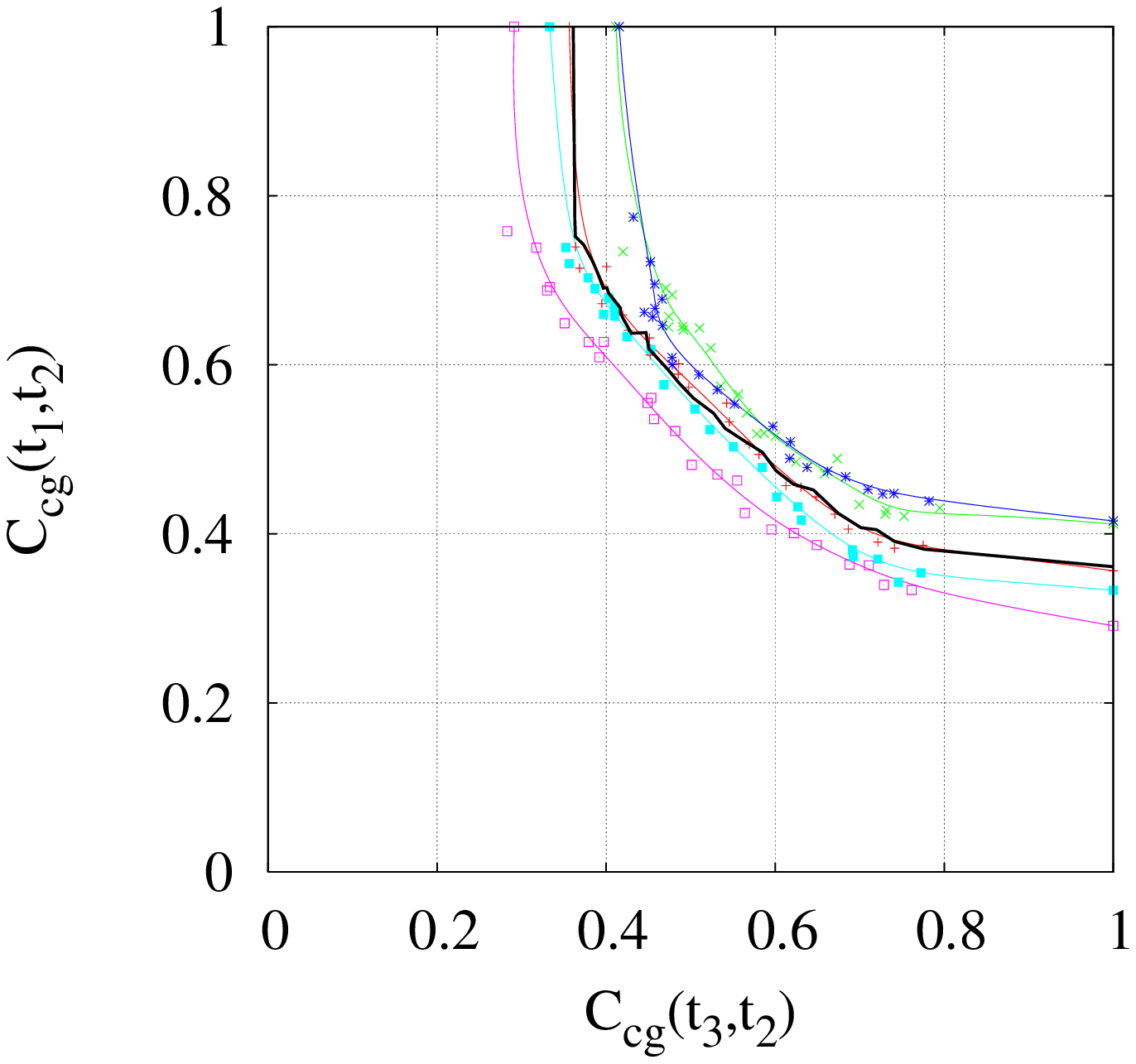}
\hspace{-1.5cm}
\includegraphics[width=8cm]{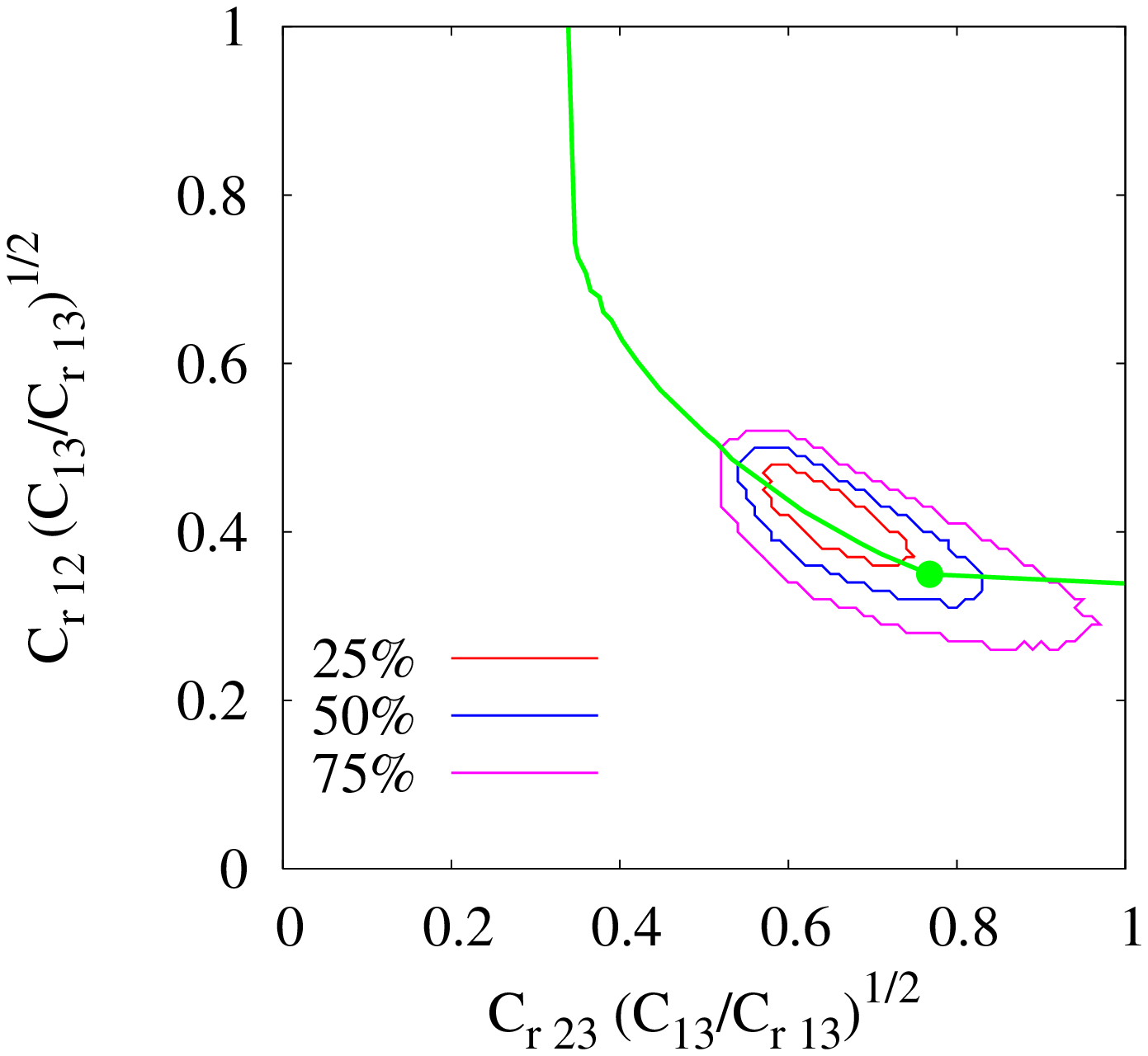}
\end{flushright}
\begin{flushright}
\includegraphics[width=8cm]{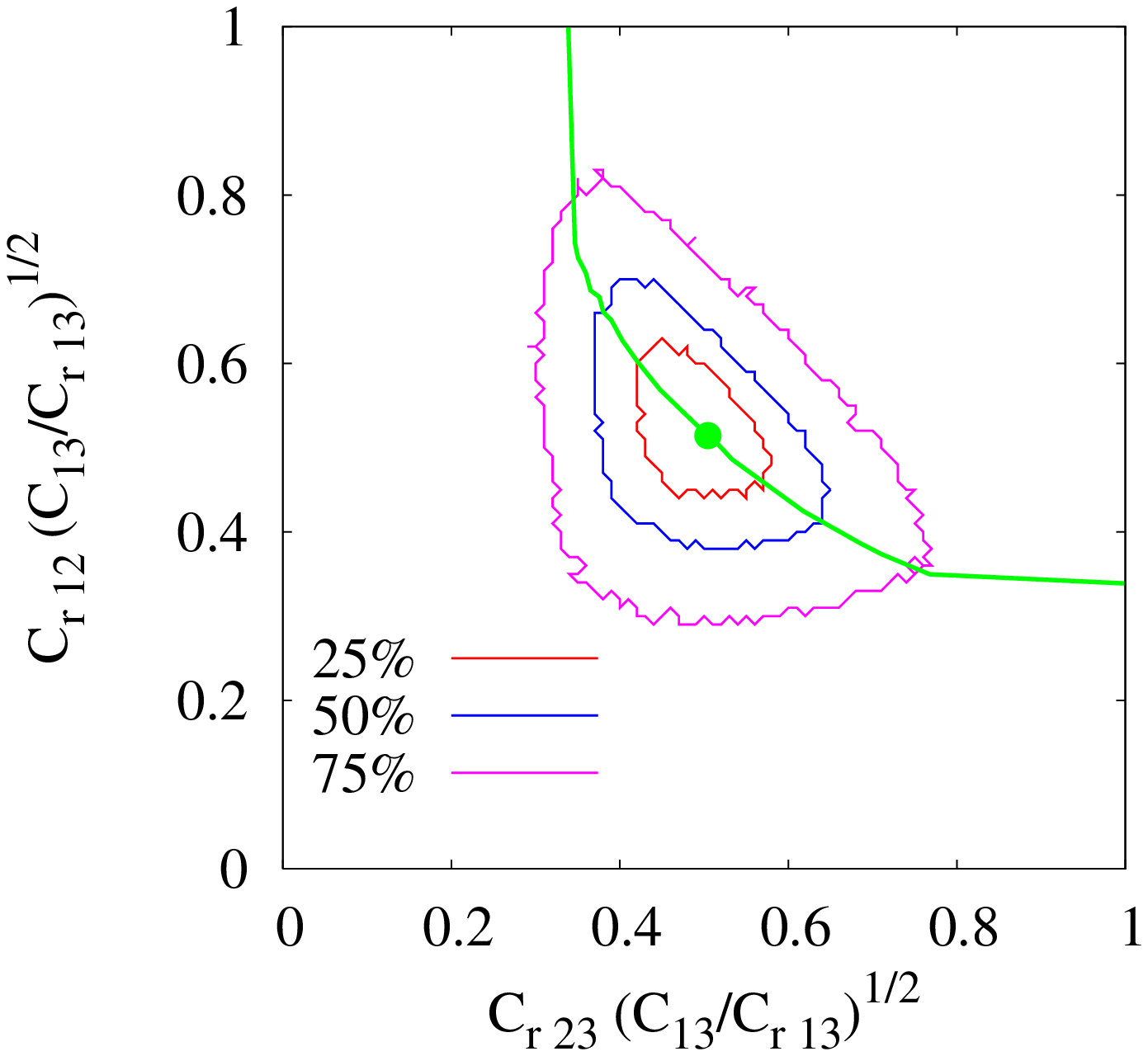}
\hspace{-1.5cm}
\includegraphics[width=8cm]{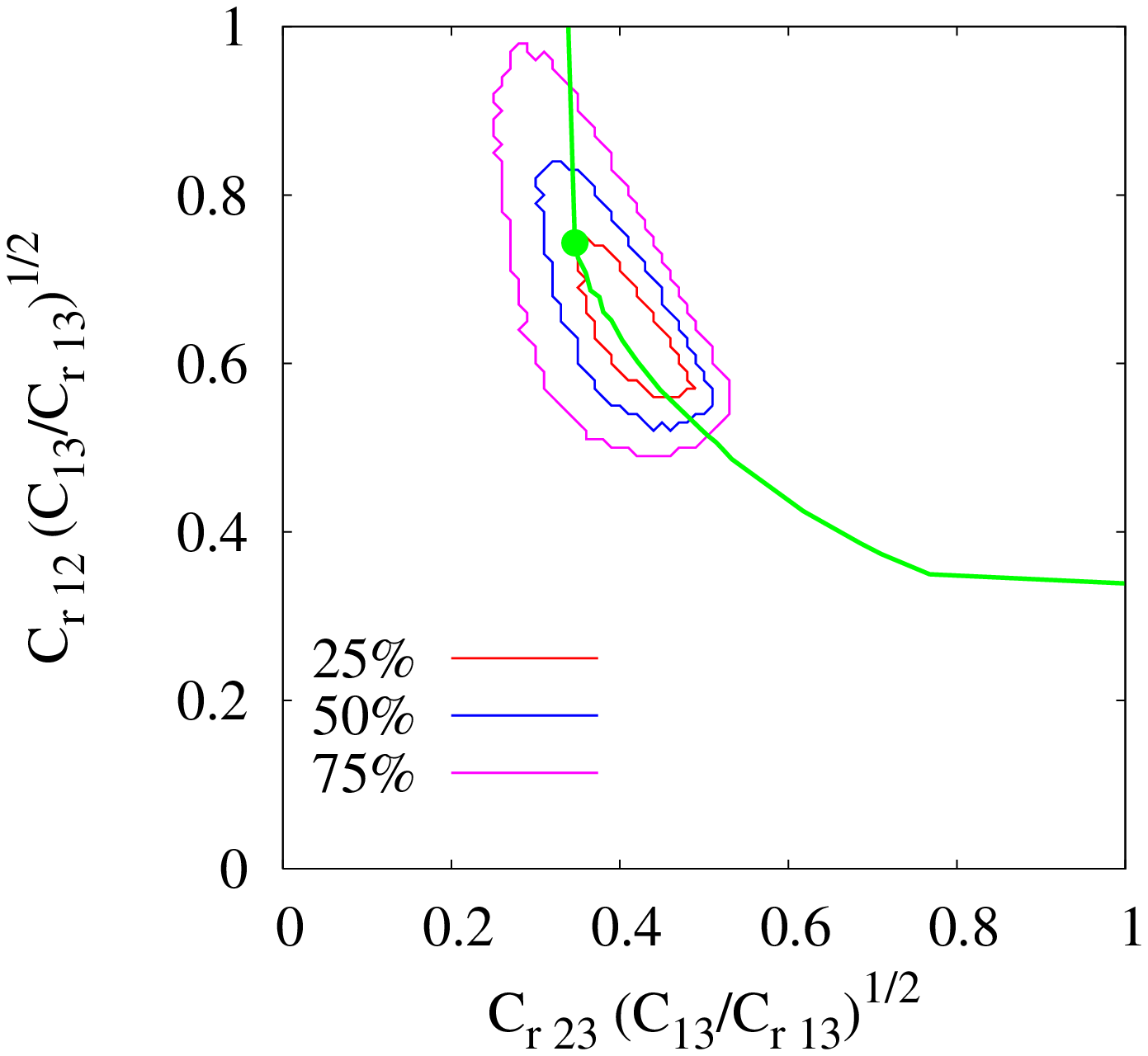}
\end{flushright}
\caption{The triangular relation in the $3d$EA model with $L=100$. Upper
left panel: the thick (black) line represents the global
$C(t_1,t_2)$ against $C(t_2,t_3)$ using $t_2$ as a parameter varying
between $t_3=5 \times 10^4$ MCs and $t_1=9 \times 10^6$ MCs,
$C(t_1,t_3)\sim 0.35$ and $q_{ea}\sim 0.8$, 
{\it cfr.} Fig.~\ref{fig:triangular}. With different points joined
with thin lines we show the triangular relations between the local
coarse-grained correlations on five randomly chosen sites on the
lattice ($\ell=30$).  Upper right panel and lower left and right
panels: $2d$ projection of the joint probability density of $C(r;
t_1,t_2)\sqrt{C(t_1,t_3)/C(r;t_1,t_3)}$ and $C(r;
t_2,t_3)\sqrt{C(t_1,t_3)/C(r;t_1,t_3)}$ at fixed three values of the
intermediate time, $t_2=1.5\times 10^5$ MCs, $8\times 10^{5}$ MCs
and $5 \times 10^6$ MCs, respectively and $\ell=10$. The global
$C(t_1,t_2)$ against $C(t_2,t_3)$ using $t_2$ as a parameter is
shown with a thick green line. The green points indicate the
location of $C(t_1,t_2)$ and $C(t_2,t_3)$ for the chosen
$t_2$'s. Each point in the scatter plot corresponds to a site,
$r$. The lines indicate the boundary surrounding 25\%, 50\% and 75\%
of the probability density.  The cloud extends mostly along the
global relation as predicted by time-reparametrization
invariance.}
\label{fig:triangular-3dEA}
\end{figure}

In Sect.~\ref{sec:global-corr}  we showed that two-time
functions with a separation of time-scales as in
eq.~(\ref{eq:additive-scaling}) and an aging scaling as in
eq.~(\ref{eq:scaling-C3dEA}) are related in a parametric way in which
times disappear.
Equation~(\ref{eq:fluct-cchi}) implies that the local (fluctuating)
two-time functions should verify the {\it same} relation
\begin{equation}
C_{ag}(r;t_1,t_3) = q_{ea} \; 
f\left\{ 
f^{-1}[C_{ag}(r;t_1,t_2)/q_{ea}] \; 
f^{-1}[C_{ag}(r;t_2,t_3)/q_{ea}] 
\right\} 
\; . 
\label{eq:triangular2} 
\end{equation} 
This is a result of the fact that time-reparametrization invariance
restricts the fluctuations to appear only in the local functions
$h(r,t)$ while the function $f$ is locked to be the global
one everywhere in the sample.  

A pictorial inspection of this relation should take into account the
fact that while the stationary decay is not expected to fluctuate,
the full aging relaxation and, in particular, the minimal value of
the local two-time functions, $C(r;t_1,t_3)$, are indeed
fluctuating quantities. The parametric construction on different
spatial regions should yield `parallel translated' curves with
respect to the global one, as displayed in
Fig.~\ref{fig:triangular}.  Fluctuations in the function $f$
would yield different functional forms in the curved part of the
parametric construction.  A more quantitative analysis can be done
by using the knowledge of $f$ that can be extracted from the
global correlation decay. Indeed, if $f$ is known, the parametric
plot $f^{-1}(C_{r12}/q_{ea})/\sqrt{f^{-1}(C_{r13}/q_{ea})}$ against 
$f^{-1}(C_{r23}/q_{ea})/\sqrt{f^{-1}(C_{r13}/q_{ea})}$ should yield a
master curve identical to the global one with different sites just
being advanced or retarded with respect to the global value.   This
is another way of stating that the sample ages in a heterogeneous
manner, with some regions being younger (other older) than the
global average. [For simplicity we used a short-hand notation,
$C_{r\mu\nu}=C(r; t_\mu,t_\nu)$ with $\mu,\nu=1,2,3$.]
If the time-reparametrization mode is indeed flat the local values
should lie all along this master curve in the aging regime.

The conclusions drawn above apply in the strict $1\ll \ell \ll \xi$
limit. In simulations and experiments $\xi$ is finite  and even
rather short. Thus, $\ell$ is forced to also be a rather small
parameter, in which case  `finite size' fluctuations in $f$ are
also expected to  exist. The claim is that the latter should scale
down to zero  faster (in $\ell$) than the fluctuations that are
related to  the zero mode. 

We have tested these claims in the non-equilibrium dynamics of the
$3d$ {\sc ea} spin-glass. The results are shown in
Fig.~\ref{fig:triangular-3dEA}. In the upper left panel we show the
global triangular relation (thick black line) as well as the local one
on four chosen sites.  The separation of time-scales is clear in the
plot. The local curves are quite parallel indeed.  In the remaining
panels in Fig.~\ref{fig:triangular-3dEA} we show the $2d$ projection
of the joint probability density (using contours that encircle
$25\%,50\%$, and $75\%$ of the probability) of the site fluctuations
in the local coarse-grained correlations at different chosen times
$t_2=1.5 \times 10^5$ MCs (upper right), $t_2=8 \times 10^5$
MCs (lower left), and $t_2=5 \times 10^6$ MCs (lower right).  Taking
advantage of the fact that $f(x) \sim x^{-1}$ we use a very
convenient normalization in which we multiply the horizontal and
vertical axes by $\sqrt{C_{13}/C_{r13}}$.  Global
time-reparametrization invariance, expressed in
eq.~(\ref{eq:triangular2}), implies that the data points should spread
{\it along} the global curve indicated with a thick green line in the
figure. Some sites could be advanced, others retarded, with respect to
the global value -- shown with a point on the green curve -- but all
should lie mainly on the same master curve.  This is indeed quite well
reproduced by the simulation data in the three cases, $C(t_1,t_2)$
close to $C(t_1,t_3)$ (upper right panel), $C(t_1,t_2)$ close to
$q_{ea}$ (lower right panel), and $C(t_1,t_2)$ far from both (lower
left panel). Most of the data points tend to follow the master curve
though some fall away from it. The reason for this is that
eq.~(\ref{eq:triangular2}) should be strictly satisfied only in the
very large coarse-graining volume limit ($\ell \gg 1$) with $\ell/\xi
\ll 1$ while we are here using $\ell=10 \stackrel{>}{\sim} \xi$, see
the discussion in Sect.~\ref{subsect:corr-length}.

Finally, notice that, in contrast to the growing correlation length
scale, there is no blatantly obvious explanation of these triangular
relations within other theoretical scenarios. These relations are
perhaps the most direct consequence of the time reparametrization
symmetry arguments.

\subsection{Normal form of the pdfs} 
 
Once we established that, due to the growth of the correlation 
length, it is necessary to use different coarse-graining lengths to
compare the pdf of local correlations evaluated at different times we
now go back to the raw data and we study the functional form of the
pdfs, {\it cfr.} Fig.~\ref{fig:rhoCr-generic}.  The reason for this
study is that the time-reparametrization invariance scenario yields
precise predictions on the functional form that one would like to
test~\cite{Ch04}.

\begin{figure} 
\centerline{ 
\includegraphics[width=8cm]{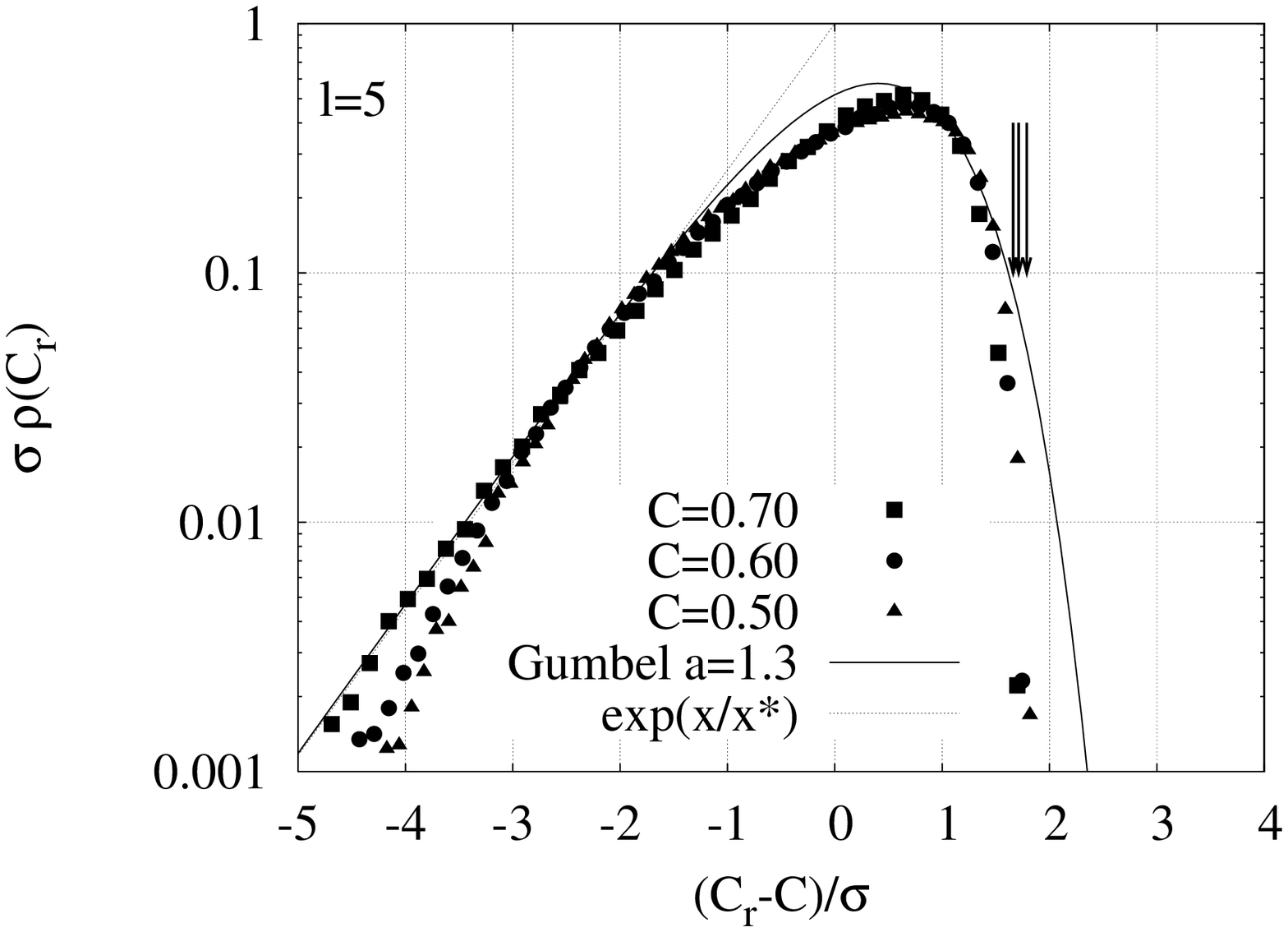} 
} 
\centerline{ 
\includegraphics[width=8cm]{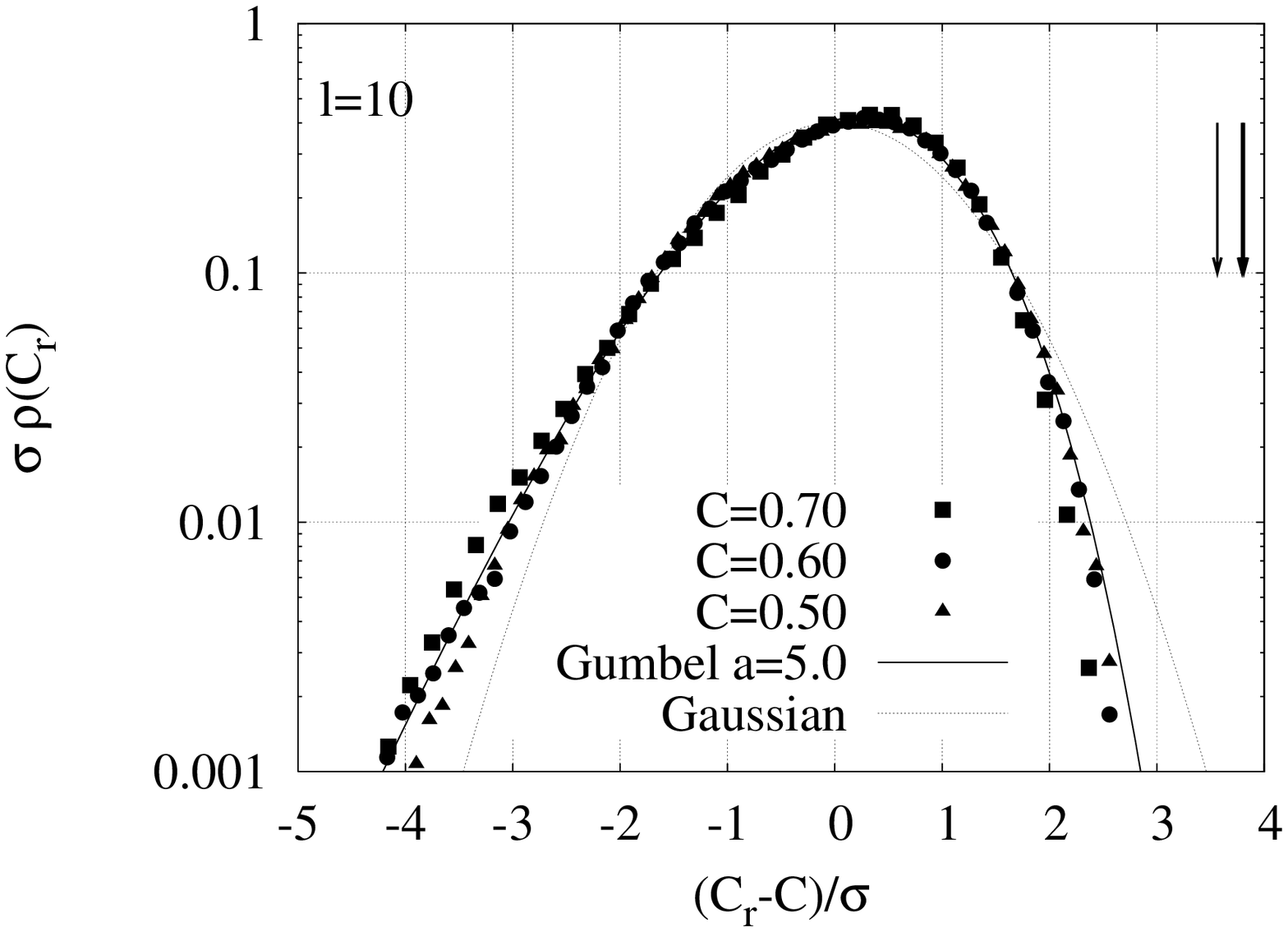} 
} 
\centerline{ 
\includegraphics[width=8cm]{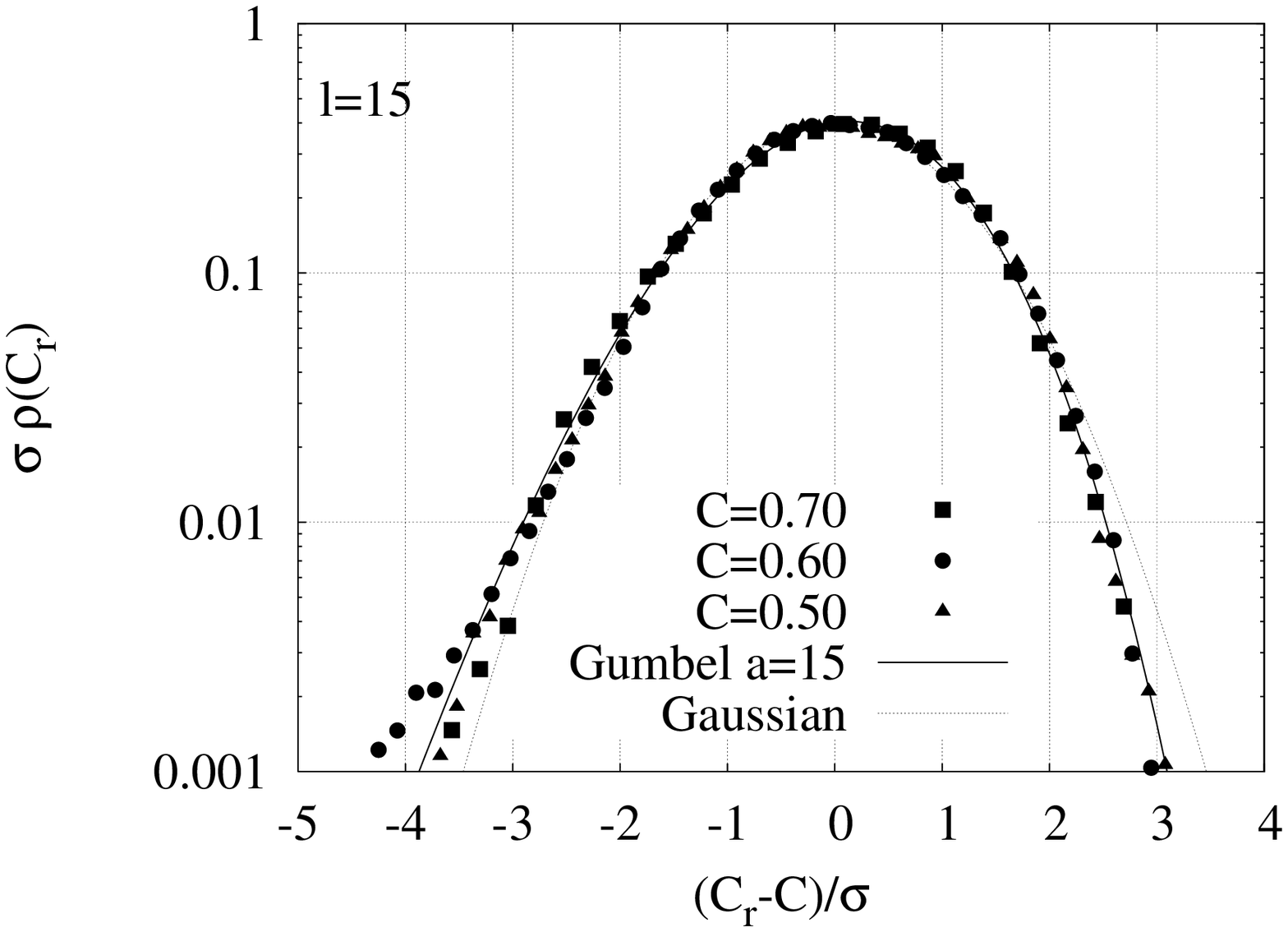} 
} 
\caption{Normal form of the pdf of local coarse-grained correlations 
at $t_w=5 \times 10^4$ and $t=1.2 \times 10^5$ ($C=0.701$), $t=3 
\times 10^5$ ($C=0.601$) and $t=10^6$ ($C=0.503$) and different values 
of the coarse-graining length $\ell=5,\; 10,\; 15$ as indicated in the 
keys. Comparison to different functional forms, in particular the 
Gumbel (solid line) and the Gaussian (dashed line) functions. The 
three arrows indicate the right-edge of the numerical distribution, 
$(1-C)/\sigma$.  The values of $\sigma$ are the following.  When the 
global correlation is $C=0.701$: $\sigma=0.180$ for $\ell= 5$, 
$\sigma=0.0785$ for $\ell=10$ and $\sigma=0.0455$ for $\ell=15$.  When 
the global correlation is $C=0.601$: $\sigma=0.233$ for $\ell= 5$, 
$\sigma=0.112$ for $\ell=10$ and $\sigma=0.0675$ for $\ell=15$.  When 
the global correlation is $C=0.503$: $\sigma=0.278$ for $\ell= 5$, 
$\sigma=0.131$ for $\ell=10$ and $\sigma=0.0772$ for $\ell=15$.  Note 
that the exponential fit of the left tail gives a quite correct description 
for $C=0.701$ and $\ell=5$ but for larger values of $\ell$ the tail 
deviates since the pdf is approaching the Gaussian.} 
\label{fig:pdf-normal} 
\end{figure}

To this end, we change variables and we plot $\sigma \rho(C_r)$
against $(C_r-C)/\sigma$ with $\sigma$ the mean-square
displacement. Note that, due to the scaling hypothesis $\sigma$ also
depends on $C$ and $\ell/\xi$.  Evidently, the transformation into the
normal form is not sufficient to eliminate the dependence on $C$ and
$\ell/\xi$ from $\rho$.  Keeping this fact in mind, we present data
for $\sigma \rho(C_r) >10^{-3}$ in Fig.~\ref{fig:pdf-normal}.

Several features of these curves deserve a careful description.
First, the abrupt end of the data on the right is due to the finite
values of $\ell$. Indeed, the probability is strictly zero for $C_r>1$
while it is notably different from zero for $C_r=1$ as soon as $\ell$
is finite and not too large.  Second, for small values of $\ell$ the
peak is located close to $q_{ea}$ while it moves to the left for
increasing $\ell$. Third, it is clear that the decay on the right of
the peak is much faster than the one on the left.  Fourth, for small
$\ell$ the dependence on $C$ appears on the tails (left and right) and
it disappears close to the peak, {\it cfr.}
Fig.~\ref{fig:rhoCr-generic}.  

One can now try to find a convenient functional form to describe the
numerical data for sufficiently large $\ell$, say of the order of
$\xi(t,t_w) \sim 5$.  In~\cite{Ch04} the pdfs of the local connected
correlations in kinetically facilitated models and the global
spin-spin correlation (\ref{eq:self-correlation}) in small size 
$3d$EA models for values of $C$ that are not too far from 
$q_{ea}$ were analysed using an extension of the Gumbel distribution
of extreme value statistics:
\begin{eqnarray} 
\Phi_a(y) & = & \frac{|\alpha| a^a}{\Gamma(a)}  
\; {\rm e}^{a\left(\alpha(y-y_0)-{\rm e}^{\alpha(y-y_0)}\right)} \,, 
\label{Gumbel-dist} 
\end{eqnarray} 
where $\Gamma(a)$ is the {\it gamma} function.  The parameters $y_0$
and $\alpha$ control the position of the center and the width of the
distribution, respectively. By fixing the parameters $\alpha =
\sqrt{\Psi'(a)}$ and $y_0 = [\log a - \Psi(a)]/\alpha$ where $\Psi(a)$
is the {\it digamma} function $\Psi(a)=\Gamma'(a)/\Gamma(a)$, one
ensures that the center is at zero and the width is unity.  The Gumbel
distribution with with {\it integer} parameter $a\geq 1$ describe the
distribution of the $a$-th largest (smallest) value in a sequence of
independent identically distributed random variables with a
probability density decaying faster than any power law.  For any $a$,
there is a choice of sign for $\alpha$ and the two signs correspond to
the Gumbel statistics of either extreme minima or extreme maxima.  If
the Gumbel is taken to its normal form by fixing the coefficient
$\alpha$, it approaches the normal form of the Gaussian for increasing
$a$.

In Fig.~\ref{fig:pdf-normal} we compare the numerical data to the
functional form (\ref{Gumbel-dist}) and the Gaussian with zero mean and unit
variance. The data on each panel are for three values of the global
correlation $C\sim 0.7$, $C\sim 0.6$ and $C\sim 0.5$ not too far from 
$q_{ea}\sim 0.8$; the difference
between panels is the length of the coarse-graining box, $\ell=5$ (a),
$\ell=10$ (b), $\ell=15$ (c). For small $\ell$ the complete data
deviate from both functional forms. There is, as expected, a systematic
dependence on $C$ that is most clear on the left tail.  An exponential
decay of the tail on the left (as found in the Gumbel form) yields a
satisfactory description of the data for each $C$ at sufficiently
small $(C_r-C)/\sigma$, as shown by the straight line.  As $\ell$
increases the data get closer to (\ref{Gumbel-dist}) on the whole
range -- note however that the numerical data are constrained to vary
between $(-1-C)/\sigma$ and $(1-C)/\sigma$ on the horizontal axis
(shown with vertical arrows) while the independent variable in the
functional form (\ref{Gumbel-dist}) can take any real value. In panels
(b) and (c) there is no systematic dependence on $C$. The value of the
parameter $a$ has been chosen to yield the best description of the
full set of data, finding that $a$ increases with increasing
$\ell$ thus approaching a Gaussian
(though it has not reached it yet for $\ell =15$).

In summary, for {\it finite} $\xi$ one identifies three functional forms of  
the pdf of local correlations, once 
put in normal form: 
\begin{itemize} 
\item 
$\ell \ll \xi$ 
Nothing special (since for small $\xi$ this means $\ell$ of the order of the 
lattice spacing).
\item 
$\ell \sim \xi$ 
Gumbel-like.
\item 
$\ell \gg \xi$ Gaussian.
\end{itemize} 
Note that in the ideal $t, \, t_w\to\infty$, and $\xi \to \infty$ limits
one  should have an interesting scaling regime  when $\ell/
\xi\to 0$, now with $\ell$ very large. 
 
Let us focus on sufficiently long coarse-graining lengths such that
the Gumbel and then Gaussian functional forms give a reasonable
description of the data with a variable $a$.  The full pdf and hence
the parameter $a$ itself depend on the two times and the
coarse-graining length $\ell$. Trading the two times by the global
correlation and the correlation length, as explained in
Sect.~\ref{sec:local}, one has $a(t,t_w,\ell) \to
a(C,\xi,\ell)$. One again, our scaling hypothesis implies
$a(C,\ell/\xi)$. Quite remarkably, the pdfs once presented in normal
form, and in particular $a$, do not show a strong dependence on
$C$ in the accessed range of values.

  Recently, Bertin and Clusel gave a simple explanation of the
ubiquity  of Gumbel-like probability distributions of 
fluctuations~\cite{Be06}. They showed that sums of infinitely many 
correlated random variables, with special correlations, are 
distributed according to a Gumbel functional form. In the case in 
which the correlation arises due to a rewriting of the maximum of a 
sequence as a sum over ordered differences, one finds the ordinary 
Gumbel pdf of extreme value statistics. For other, more general 
correlations one finds that the sum is distributed according to the 
Gumbel functional form with a continuous parameter $a$ (even other 
functions are possible). In our problem, the local coarse-grained
correlations are,  by construction, sums of correlated random
variables. The fact that we  find a functional form that resembles a
generalized Gumbel  distribution, with a parameter that depends on
$\ell$ and the value of  the global correlation, is then consistent
with the observation of Bertin  and Clusel.     

From our point of view, these results are interesting and support the
conjecture that the development of global time-reparametrization
invariance is the mechanism by which dynamic fluctuations generate in
glassy dynamics.  Indeed, the phenomenological effective action for
local fluctuations proposed in~\cite{Ch04} suggested by this symmetry
captures this functional form when $C\stackrel{<}{\sim} q_{ea}$. 

\section{Conclusions} 
\label{sec:conclusions} 
 
The peculiar dynamics of glassy systems, leading to an asymptotically
divergent correlation length, implies that highly non-trivial
fluctuations will still be present at longer and longer length scales
as times increase. Local fluctuations can be captured by studying
coarse-grained local two-time correlations, which provide a measure of
local aging in the system, while the global measure of aging is
obtained by simply taking the coarse-graining length to be the bulk
system size $L$. The presence of a growing length scale affects the
distributions of local correlations, in that larger and larger regions
in space become correlated, and the local average over a
coarse-graining region will not be simply Gaussian-distributed if the
coarse-graining length $\ell$ is not much larger than the two-time
correlation length $\xi(t,t_w)$.

In this paper we studied the effect of such growing length scale on
the scaling of the distributions of local correlations in the $3d$EA
model. We found that the pdfs of local coarse-grained correlations
$\rho(C_r;t,t_w,\ell,L)$ can be scaled into universal curves that
depend on the four parameters $t,t_w,\ell,L$ only through three
scaling variables: the value of the global correlation $C(t,t_w)$, and
the ratios between the length scales and the correlation length,
$\ell/\xi(t,t_w)$ and $\xi(t,t_w)/L$. (As we stressed in the paper,
the lengths $\ell$, $L$ and $\xi$ are already dimensionless as they
are measured in units of the lattice spacing, and hence the scaling as
a function of the correlation length is a scaling assumption, and does
not follow from trivial dimensionless analysis.)
The influence of a `measuring' length, in this
case the total length, was also stressed in the study of roughness
fluctuations of elastic manifolds~\cite{elastic2}.  

Because the shape of the pdfs of local correlations depend on the
correlation length and the coarse-graining length, one can use the
form of the pdfs to extract the growing length scale.  The values of
the correlation length so obtained are comparable to those acquired
directly from four-point correlation functions.  Morever, we analysed
the evolution of the functional form of the pdf of local correlations
with the coarse-graining length $\ell$.  By comparing to a Gumbel-like
functional form we studied the dependence of its characterising
parameter, $a$, with the coarse-graining volume and we linked these
observations to recent generic results on sums of correlated random
variables~\cite{Be06}.
 
The universal scalings for the pdfs of local correlations for the
$3d$EA model should be applicable to other glassy systems as well, if
there is universality in glassy behavior among systems that share
common features such as aging behavior and a growing dynamical length
scale. Indeed, progress in this direction has been already made
in~\cite{Castillo1}, where the pdfs for local correlations in the
glassy phase of Lennard-Jones binary mixtures were shown to {\it
approximately} collapse when grouped by fixing the value of the global
correlation and in \cite{Ch04} where kinetically constrained lattice
models were analysed.  According to our findings 
 the collapse in structural glasses should become even sharper if
the coarse graining box sizes are scaled 
so as to account for the growing dynamical length scale.  A similar
phenomena, though stationary in time and with a length-scale
controlled by the distance from `criticality' in temperature, are
expected to appear in super-cooled liquids too~\cite{Ludovic}.

The existence of a growing correlation length can be explained within
the time-reparametrization invariance scenario~\cite{Chcu} but it
could also be associated to other theoretical proposals such as
theories based on the mode-coupling approach or random first order
scenario~\cite{MCT} and its refinement including the effect of
entropic droplets~\cite{Wolynes}, dynamical criticality controlled by
a zero-temperature critical point~\cite{Garrahan-etal} and frustration
limited domains~\cite{Tarjus-etal}, at least in the super-cooled
liquid. We then presented an analysis of the local coarse-grained
correlations that constitutes, to our understanding, a much more
stringent test of the time-reparametrization invariance scenario. This
consists in comparing the local and global triangular relations for
correlations measured at three times. Our results are consistent with
the concrete predictions of the time-reparametrization invariance
approach. Together with the tests of the local fluctuation-dissipation
deviations~\cite{Ca03} these are the most accurate tests of the
time-reparametrization invariant theory.

\vspace{2cm} 
\noindent\underline{Acknowledgements} 
\vspace{1cm} 
 
LFC is a member of Institut Universitaire de France. We thank
H. Castillo, G. Fabricius,  T. Grigera, J. L. Iguain, H. Makse,
D. Stariolo and L. Valluzzi for very useful discussions. We wish to
especially thank L. Berthier, S. Franz and an anonymous referee for their very
useful comments on the contents and presentation of our results, 
and C. Aron for assistance in the preparation of the figures. LFC
thanks the Newton Institute at the University of Cambridge, UK, and
the Universidad Nacional de Mar del Plata, Argentina, and LDCJ the LPTHE
at the Universit\'e Pierre et Marie Curie -- Paris VI, France, for
hospitality during the preparation of this work. This work is
supported in part by the NSF Grants DMR-0305482 and DMR-0403997 (CC).

\end{document}